\documentclass[aps,superscriptaddress,amsmath,amssymb,floatfix,twocolumn,showpacs,amsfonts,longbibliography,notitlepage,superscriptaddress]{revtex4-1}
\usepackage{times}
\usepackage[varg]{txfonts}
\usepackage{textcomp}
\usepackage{graphicx}
\usepackage{subfigure}
\usepackage{color}
\usepackage[colorlinks=true,citecolor=blue,urlcolor=blue,linkcolor=blue,hyperindex]{hyperref}
\usepackage{braket}
\usepackage{float}
\usepackage{overpic}
\usepackage{bm}
\usepackage{array}
\usepackage{multirow}
\usepackage{booktabs}
\usepackage{appendix}
\usepackage{tabularx}

\makeatletter
\newcommand{\fmarki}{*}
\newcommand{\fmarkii}{\ensuremath{\dagger}}
\newcommand{\fmarkiii}{\ensuremath{\ddagger}}
\newcommand{\fmarkiv}{\ensuremath{\mathsection}}
\newcommand{\fmarkv}{\ensuremath{\mathparagraph}}
\newcommand{\fmarkvi}{\ensuremath{\|}}
\newcommand{\fmarkvii}{**}
\newcommand{\fmarkviii}{\ensuremath{\dagger\dagger}}
\newcommand{\fmarkix}{\ensuremath{\ddagger\ddagger}}

\def\@fnsymbol#1{{\ifcase#1\or \fmarki\or \fmarkii\or \fmarkiii\or \fmarkiv\or \fmarkv\or \fmarkvi\or \fmarkvii\or \fmarkviii\or \fmarkix \else\@ctrerr\fi}}
\makeatother

\renewcommand{\fmarki}{\ensuremath{\dagger}}
\renewcommand{\fmarkii}{*}
\renewcommand{\fmarkiii}{*}
\renewcommand{\fmarkiv}{*}
\renewcommand{\fmarkv}{x$_5$}
\renewcommand{\fmarkix}{z$_9$}

\begin{document}
\title{Spin density wave in the bilayered nickelate La$_3$Ni$_2$O$_{7-\delta}$ at ambient pressure}

\author{Xiao-Sheng Ni}
\affiliation{Guangdong Provincial Key Laboratory of Magnetoelectric Physics and Devices, State Key Laboratory of Optoelectronic Materials and Technologies, Center for Neutron Science and Technology, School of Physics, Sun Yat-Sen University, Guangzhou, 510275, China}
\thanks{}
\author{Yuyang Ji}
\affiliation{Key Laboratory of Quantum Information, University of Science and
	Technology of China, Hefei, 230026, People's Republic of China}
\author{Lixin He}
\affiliation{Key Laboratory of Quantum Information, University of Science and
	Technology of China, Institute of Artificial Intelligence, Hefei Comprehensive National Science Center, Hefei, 230026, People's Republic of China }
\author{Tao Xie}
\author{Dao-Xin Yao}
\author{Meng Wang}
\author{Kun Cao}
\email{caok7@mail.sysu.edu.cn}
\affiliation{Guangdong Provincial Key Laboratory of Magnetoelectric Physics and Devices, State Key Laboratory of Optoelectronic Materials and Technologies, Center for Neutron Science and Technology, School of Physics, Sun Yat-Sen University, Guangzhou, 510275, China}

\date{\today}

\maketitle

\newpage

{\bf The recent discovery of high-temperature superconductivity in high-pressurized La$_3$Ni$_2$O$_{7-\delta}$ has garnered significant attention. Using density functional theory, we investigate the magnetic properties of La$_3$Ni$_2$O$_{7-\delta}$ at ambient pressure. Our calculations suggest that with $\delta=0$, the double spin stripe phase is favored as the magnetic ground state. Oxygen vacancies may effectively turn nearest Ni spins into \textit{charge} sites. Consequently, with moderate $\delta$ values, our theoretical magnetic ground state exhibits characteristics of both double spin stripe and spin-charge stripe configurations, providing a natural explanation to reconcile the seemingly contradictory experimental findings that suggest both the configurations as candidates for the spin-density-wave phase. With higher $\delta$ values, we anticipate the ground state to become a spin-glass-like noncollinear magnetic phase with only short-range order. The oxygen vacancies are expected to significantly impact the magnetic excitations and the transition temperatures $T_{SDW}$. Notably, the magnetic ordering also induces concomitant charge ordering and orbital ordering, driven by spin-lattice coupling under the low symmetry magnetic order.
We further offer a plausible explanation for the experimental observations that the measured $T_{SDW}$ appears insensitive to the variation of samples and the lack of direct evidence for long-range magnetic ordering.}

~\\
\noindent{\bf\large Introduction}\\
The recent discovery of novel superconductivity with \(T_c\) up to 80 K in highly pressurized La$_3$Ni$_2$O$_{7-\delta}$ has sparked extensive experimental and theoretical investigations~\cite{20, 21, zhang2024high, wang2024pressure, zhang2023effects, zhou2023evidence,luo2023bilayer,yang2024orbital,luo2024high,kang2023infinite}. While current theoretical studies primarily focus on the pairing mechanisms of its high-pressure phase~\cite{PhysRevLett.131.206501, electronic2, electronic3, electronic4, electronic5, electronic6, pair3, pair4, pair5, pair6, pair7, pair1, pair2,geisler2024structural,liu2024emergence}, 
understanding the magnetic ground state of La$_3$Ni$_2$O$_{7-\delta}$ at low pressure is essential to elucidate the correlation between magnetism and the high-\(T_c\) superconductivity. For La$_3$Ni$_2$O$_{7-\delta}$ at ambient pressure, resistivity measurements have suggested a possible spin density wave (SDW) phase below a transition temperature \(T_{SDW} \sim 150\) K~\cite{liu2023evidence}. However, early nuclear magnetic resonance (NMR) experiments mainly find evidence for charge ordering in La$_3$Ni$_2$O$_{7-\delta}$, although the existence of spin density wave is not completely ruled out~\cite{kakoi2023multiband}. More recently, resonant inelastic x-ray scattering (RIXS) experiments measuring the magnetic excitations of La$_3$Ni$_2$O$_{7-\delta}$ single crystals clearly indicate the presence of in-plane magnetic correlations with \(Q = (0.25, 0.25)\) below 150 K, suggesting possible spin orders with double spin stripe (DSS) or spin-charge stripe (SCS) configurations~\cite{chen2024electronic, dan2024spin}. The presence of the DSS order is also weakly evidenced by more recent NMR data~\cite{dan2024spin} and inelastic neutron scattering data~\cite{xie2024neutron}. Conversely, the SCS is favored as the SDW phase, qualitatively consistent with a field distribution with both high and low strengths observed in a positive muon spin relaxation ($\mu^+$ SR) study of polycrystalline La$_3$Ni$_2$O$_{7-\delta}$~\cite{PhysRevLett.132.256503}.

Theoretical studies on the magnetic properties of La$_{3}$Ni$_{2}$O$_{7}$ have primarily focused on determining the magnetic ground state by comparing energies of various magnetic configurations from density functional theory (DFT) calculations~\cite{yi2024antiferromagnetic, PhysRevB.108.L180510, zhang2024structural}. Yi $et\ al.$~\cite{yi2024antiferromagnetic} identified the A-type antiferromagnetic (AFM) configuration as the magnetic ground state. Chen $et\ al.$~\cite{zhou2024electronic}, based on experimental RIXS spectra, reproduced the observed magnon dispersion using a simplified $J_1$-$J_2$-$J_3$ Heisenberg model. However, current theoretical studies on the magnetism of La$_{3}$Ni$_{2}$O$_{7}$ cannot fully explain the experimental observations. Notably, oxygen vacancies are commonly present in experimental samples of La$_{3}$Ni$_{2}$O$_{7}$. While the impact of oxygen vacancies on the electronic structure and superconductivity has been investigated based on DFT calculations~\cite{PhysRevB.83.245128, yi2024antiferromagnetic, PhysRevB.109.205156}, their effect on the magnetism of La$_{3}$Ni$_{2}$O$_{7}$ remains elusive.

In this work, we investigate the magnetic properties of La$_3$Ni$_2$O$_{7-\delta}$ at ambient pressure using DFT calculations. For $\delta = 0$, our results indicate that the DSS phase, characterized by a (0.25, 0.25) in-plane modulation vector, is the favored magnetic ground state. The presence of oxygen vacancies leads to the vanishing of Ni magnetic moments nearest to the vacancy sites, effectively creating {\it charge} sites. For moderate $\delta$ values, our theoretical SDW phase exhibits features of both DSS and SCS configurations, reconciling the seemingly contradictory experimental findings that suggest both DSS and SCS as candidates for the SDW phase. At higher oxygen vacancy concentrations, such as $\delta = 0.5$, we predict a short-range ordered ground state with spin-glass-like magnetic structures, disrupting the (0.25, 0.25) modulation vector. Oxygen vacancies also reduce the \(T_{SDW}\) due to the dilution of dominant exchange interactions. Notably, the magnetic ordering naturally brings concurrent charge ordering and orbital ordering, due to the symmetry lowering faciliated by spin-lattice coupling. Furthermore, our findings indicate that a random distribution of oxygen vacancies may need to be considered to achieve spin wave spectra consistent with experimental data. We also offer a plausible explanation for experimental observations, such as the insensitivity of the measured $T_{SDW}$ to different samples and the lack of direct evidence for long-range magnetic ordering.

\begin{figure}
	\includegraphics[scale=0.11]{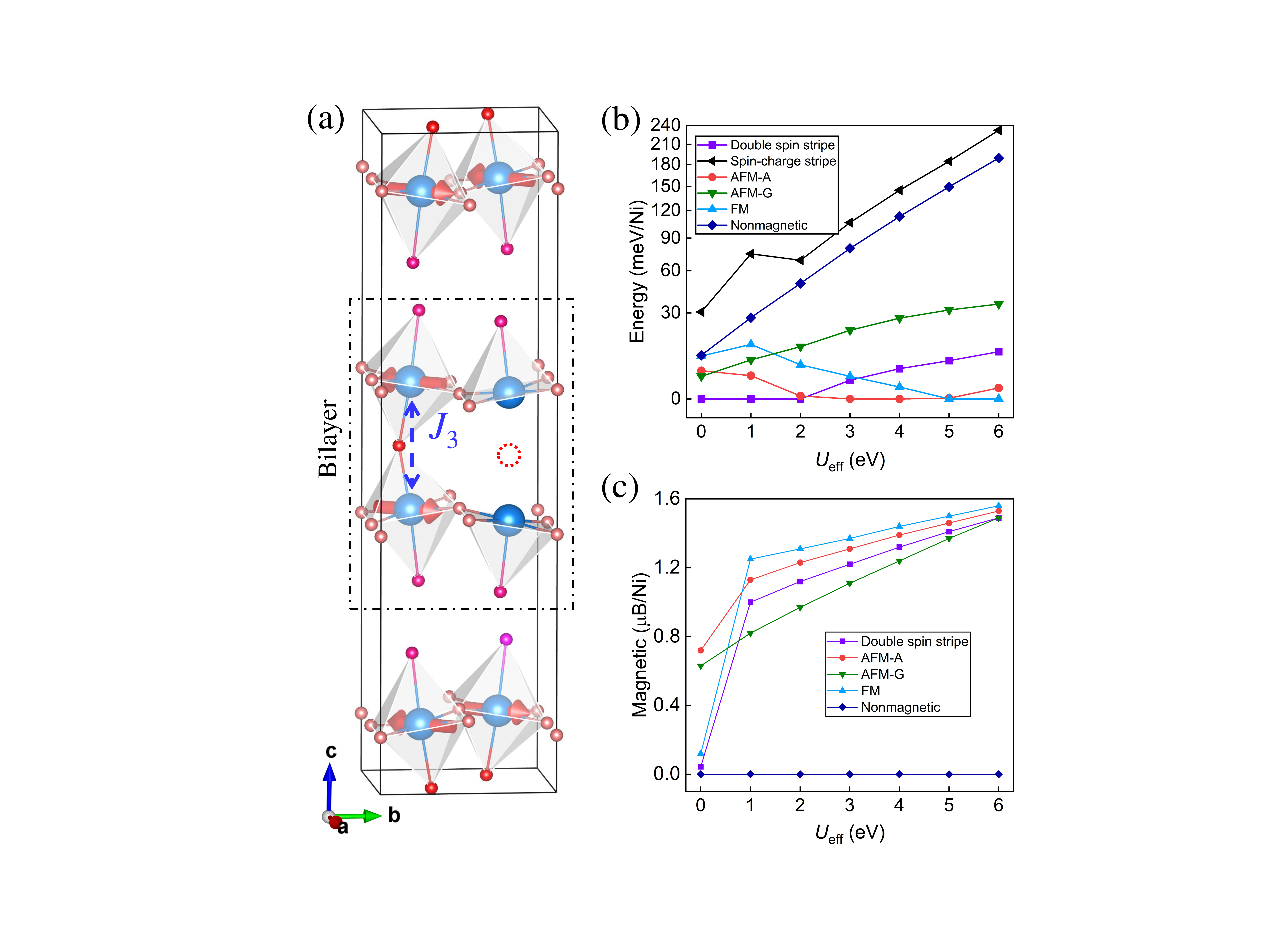}\caption{\label{fig:tu1}\textbf{Results of DFT + $U_{\text{eff}}$ and inner-apical oxygen vacancy in La$_3$Ni$_2$O$_{7-\delta}$}. (a) The crystal structure of La$_3$Ni$_2$O$_{7-\delta}$. The Ni atoms, outer-apical oxygen, inner-apical oxygen, and in-plane oxygen are represented by blue, rosy, red, and pink spheres, respectively. The red arrows denote spins. The dashed red circle indicates an inner-apical oxygen vacancy.The calculated (b) total energies and (c) average magnetic moments with different $U_{\text{eff}}$ in La$_3$Ni$_2$O$_7$. }
\end{figure}

~\\
\noindent {\bf\large RESULTS}\\
\noindent {\bf\large Crystal structure}\\
At ambient pressure, bulk La$_3$Ni$_2$O$_7$ crystallizes in an orthorhombic structure with the space group $Amam$ and lattice parameters $a = 5.393$ \AA, $b = 5.448$ \AA, and $c = 20.518$ \AA~\cite{20}. As illustrated in Fig.~\ref{fig:tu1}(a), each unit cell of La$_3$Ni$_2$O$_7$ comprises two Ni-O bilayers, with each Ni atom coordinated by six O atoms, forming NiO$_{6}$ octahedra. The oxygen atoms occupy four inequivalent sites: two within each Ni-O layer, and the other two at the inner-apical and outer-apical positions, respectively.

\begin{figure}
	\includegraphics[scale=0.10]{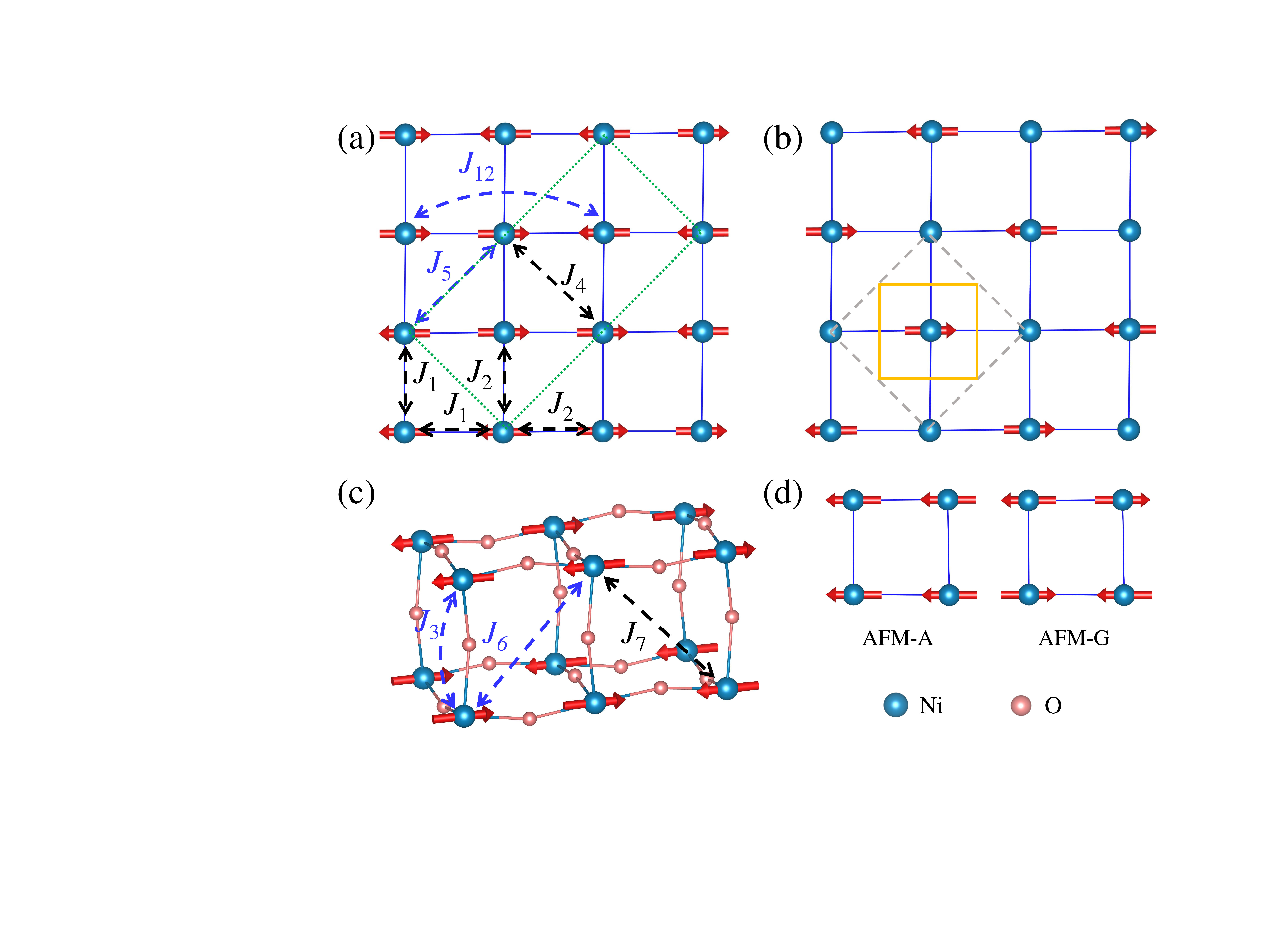}\caption{\label{fig:tu2}\textbf{Different magnetic configurations and exchange paths in La$_3$Ni$_2$O$_7$}. (a) The DSS configuration, with the exchange paths represented by dashed lines with double arrows. The green box represents a magnetic unit cell in the DSS configuration. Red arrows represent spin up and down.(b) The SCS configuration. The yellow box and the dashed box denote a unit cell in real space and the corresponding first Brillouin zone, respectively. (c) Exchange paths for $J$’s are represented by dashed lines with arrows, where black color indicates ferromagnetic interactions and blue color indicates antiferromagnetic interactions. Ni and O atoms are illustrated as blue and pink spheres, respectively. (d) Schematic representation of different magnetic configurations. }
\end{figure}

~\\
\noindent {\bf\large Magnetic order}\\
Magnetic properties from DFT + $U$ calculations are typically sensitive to the $U$ values. 
Therefore, we first perform total energy calculations for several representative magnetic configurations across a range of $U$ values using the DFT+ $U$ method~\cite{dudarev1998electron}.
We consider the ferromagnetic (FM), A-type AFM (AFM-A), G-type AFM (AFM-G), DSS, and SCS configurations as illustrated in Fig.~\ref{fig:tu2}(a)-(b) and (d). The DSS corresponds to an up-up-down-down stripe arrangement along each in-plane lattice vector in the tetragonal lattice setting, while the SCS can then be constructed by alternately replacing half of the local spins in the DSS by spinless charge sites. To be consistent with experimental literatures~\cite{chen2024electronic, dan2024spin, xie2024neutron}, we adopt the DSS and SCS to describe these two spin configurations throughout our paper.
Our calculated energies indicate that the DSS configuration has the lowest energy among the selected configurations when $U_{\text{eff}}$ is no larger than 2 eV (see Fig.~\ref{fig:tu1}(b)), consistent with experimental findings \cite{dan2024spin, xie2024neutron}. In the $U_{\text{eff}}$ range of 3 to 4 eV, the AFM-A configuration emerges as the magnetic ground state. Conversely, for $U_{\text{eff}}$ values exceeding 5 eV, the ground state transitions to FM. The only distinction between the AFM-A and FM configurations lies in the interlayer magnetic correlations, suggesting a shift from AFM to FM interlayer interactions with increasing $U_{\text{eff}}$(see Fig.~\ref{fig:tu1}(b)). Although the SCS has been proposed to fit the experimental magnetic excitation spectra well, its calculated energy consistently exceeds that of the DSS across all $U_{\text{eff}}$ values, favoring the DSS as the SDW phase. 
The results from our DFT$+U$ calculations are further supported by our HSE06 calculations (see Supplementary Note 3 Table SII). Moreover, the calculated magnetic moments increase with larger $U_{\text{eff}}$ values due to the increased localization of Ni $3d$ electrons. Given that experimental measurements report relatively weak magnetic moments, with average values of 0.08 $\mu_B$/Ni~\cite{dan2024spin} and 0.55 $\mu_B$/Ni~\cite{khasanov2024pressure}, we primarily present results with $U_{\text{eff}} = 1$ eV unless stated otherwise. 
Although larger \(U\) values (\(U\) $\sim$ 2 - 4 eV) are usually employed to study electronic structures of La\(_3\)Ni\(_2\)O\(_7\) ~\cite{PhysRevLett.131.206501, labollita2023electronic}, especially for the case under high pressure, our calculations indicate that a smaller \(U\) value may be more suitable for the description of its magnetic properties, which is also consistent with the findings of two recent theoretical studies on the SDW of La\(_3\)Ni\(_2\)O\(_7\) ~\cite{labollita2024assessing,zhang2024emergent}.
It is worth noting that, although all Ni atoms occupy symmetrically equivalent sites in the space group $Amam$, the symmetry lowering induced by the DSS phase causes the calculated magnetic moments to split into two groups, with amplitudes of 1.2 $\mu_B$ and 0.7 $\mu_B$, respectively(see Supplementary Note 1 TABLE SI and Fig.~\ref{fig:tu1.5}).

To further explore the magnetic properties, we employ the classical Heisenberg model to describe the magnetic interactions in La$_3$Ni$_2$O$_7$, represented by the Hamiltonian,
\begin{equation}
	\textit{H} = \sum_{i<j}J_{ij}\textbf{S}_i \cdot \textbf{S}_j
\end{equation}
where \( J_{ij} \) are the Heisenberg exchange interactions between Ni spins \(\boldsymbol{S}_i\) and \(\boldsymbol{S}_j\).
To adequately capture the relevant magnetic interactions, we consider exchange interactions with Ni-Ni bond lengths up to 8 \AA. In total, there are 12 inequivalent types of \( J \)'s, among which \( J_1 \), \( J_2 \), \( J_4 \), \( J_5 \), and \( J_{12} \) are interactions within each Ni-O layer, while \( J_3 \) and \( J_6 \) - \( J_9 \) are interlayer interactions. \( J_{10} \) and \( J_{11} \) represent interactions between bilayers. The major exchange interactions are illustrated in Fig.~\ref{fig:tu2}(a) and (c), with calculated results presented in Table~\ref{tab1}.
Within each Ni-O layer, the FM interactions \( J_1 \) and \( J_4 \), and the AFM interactions \( J_5 \) and \( J_{12} \), favor the formation of an in-plane magnetic structure with a wavevector of (0.25, 0.25). The FM interaction \( J_2 \) impedes the formation of this magnetic phase, introducing magnetic frustration. However, this frustration is relatively weak, thus allowing the (0.25, 0.25) magnetic phase to persist. On the other hand, the interlayer interactions are predominantly governed by the strong AFM interaction \( J_3 \), which is an order of magnitude larger than the other $J$'s. It is worth noting that the DSS and SCS only refer to in-plane spin arrangement, which is not directly affected by the strong \( J_3 \). However, in practice, antiferromagnetic interlayer coupling is implied in the DSS and SCS\cite{zhou2024electronic}. In this sense, the interlayer coupling affects the formation of the DSS and SCS, by enforcing an AFM arrangement between the Ni atom layers within each bilayer. In contrast, the inter-bilayer interactions \( J_{10} \) and \( J_{11} \) are negligibly small. These results are consistent with the exchange interactions extracted from experimental RIXS data~\cite{zhou2024electronic}.

\begin{table*}[t!]
	\caption{ \label{tab1} Calculated Heisenberg exchange interactions with $\delta$ = 0. The values of  \(\boldsymbol{S}_i\) are all normalized to 1. Positive and negative values represent AFM and FM interactions, respectively.}
	\begin{ruledtabular}
		\begin{tabular}{ccccccccccccccc}\\[-2.0ex]
			& La$_{3}$Ni$_{2}$O$_{7}$ & $ J_{1}$ & $ J_{2}$ & $J_{3}$ & $ J_{4}$  & $ J_{5}$ & $J_{6}$ & $ J_{7}$ & $ J_{8}$ & $ J_{9}$ &$ J_{10}$ &$ J_{11}$ &$ J_{12}$ \\
			\hline  \\[-1.0ex]
			&  Bond length (\AA)&3.806& 3.859 & 3.945 & 5.393 & 5.449 & 5.482 & 5.520 &6.682  & 6.727 & 6.866 & 6.877 & 7.666  \\
			 & Value (meV)&-6.63
			& -3.09 & 49.55  & -2.80  & 2.09  & 1.17   & -0.10  & 0.67   & -2.63   & 0.01  & 0.09   & 6.75   \\
		\end{tabular}
	\end{ruledtabular}
\end{table*}

 
Based on the calculated \( J \)'s values, we explore the magnetic phase diagrams using a replica-exchange Monte Carlo (MC) method~\cite{swendsen1986replica,mc}. Our MC simulations confirm the DSS as the ground state, consistent with our total energy calculations. However, the simulated \( T_{SDW} \) is $\sim$ 232 K, which is significantly higher than the experimental values~\cite{liu2023evidence,kakoi2023multiband,chen2024electronic, dan2024spin}. 
This discrepancy could be attributed to limitations in our DFT + \( U \) calculations, which may not accurately capture all relevant magnetic interactions. 
Additionally, we find that oxygen vacancies may also play an important role in the magnetic properties, which we will discuss further in the subsequent sections.

\begin{figure}
	\includegraphics[scale=0.15]{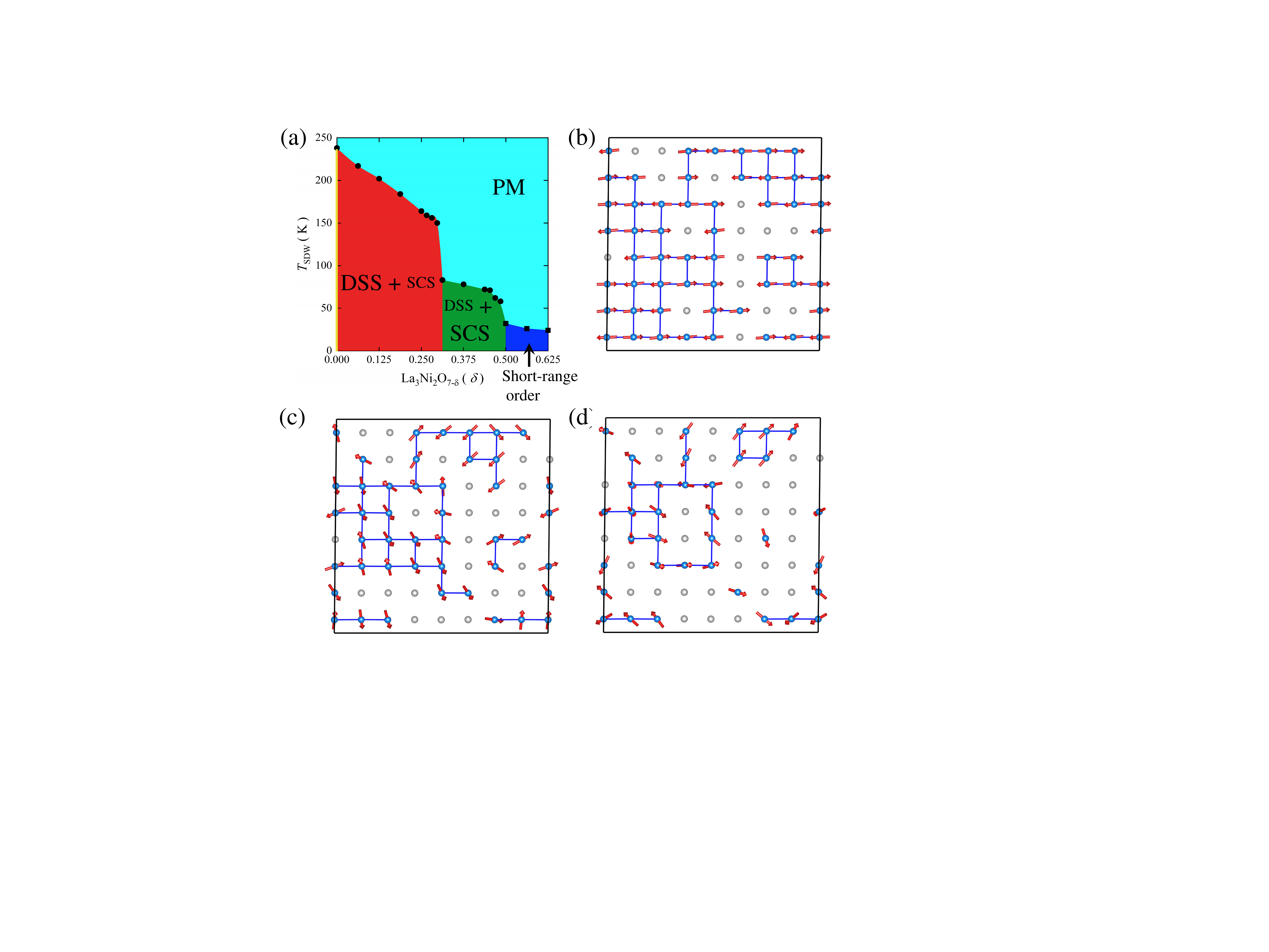}\caption{\label{fig:tu3}\textbf{Phase diagram of La$_3$Ni$_2$O$_{7-\delta}$}. (a) Phase diagram of La$_3$Ni$_2$O$_{7-\delta}$ obtained from MC simulations, where PM stands for paramagnetic. 
		(b)-(d) Sketches of typical magnetic ground states from MC simulations with $\delta$ of 0.25, 0.375, and 0.5, respectively. Red arrows represent spin up and down.}
\end{figure}

~\\
\noindent {\bf\large Oxygen vacancy}\\
We conduct first-principles calculations to investigate the oxygen vacancy configurations for three types of oxygen positions: inner-apical oxygen, in-plane oxygen, and outer-apical oxygen (see Fig.~\ref{fig:tu1}(a)). Our calculations reveal that the inner-apical oxygen vacancy is the most stable, with significantly lower formation energies compared to the other two types of vacancies. This finding suggests that the inner-apical oxygen vacancies are more likely to form during practical material synthesis, consistent with experimental observations reported in Ref.~\cite{dong2023visualization}. Therefore, we only consider the inner-apical oxygen vacancies, implemented by removing the corresponding number of inner-apical oxygen atoms in a La$_3$Ni$_2$O$_7$ supercell (see Supplementary Note 4). In experimental samples, the $\delta$ values are found to typically range from 0 to 0.5~\cite{gao2024la3ni2o6,dong2023visualization,PhysRevLett.132.256503,dan2024spin,khasanov2024pressure}. To reduce computational cost from the supercell approach, we mainly perform DFT calculations with $\delta = 0.25$ and $\delta = 0.5$ using a 48-atom supercell. We then investigate the effect of vacancy concentrations on the magnetism and find that the magnetic moments of the Ni atoms nearest to the oxygen vacancies diminish to nearly zero ($\sim$0.03 \(\mu_B\)), effectively forming {\it charge} sites. As shown in Fig.~\ref{fig:tu1}(a), two charge sites can be produced by removing one oxygen atom. As a result, the presence of charge sites leads to a noticeable reduction in the average magnetic moment, which may contribute to the experimentally observed small magnetic moments of Ni~\cite{dan2024spin, khasanov2024pressure, PhysRevLett.132.256503}.


We then recompute the exchange interactions of La$_3$Ni$_2$O$_{7-\delta}$ for $\delta = 0.25$ and $\delta = 0.5$, under which the original 12 $J$'s split into 48 $J$'s and 16 $J$'s, respectively (see Tables SIII and SV in Supplementary Note 4). Unsurprisingly, with oxygen vacancies, the $J$'s involving the introduced {\it charge} sites are among the most affected, with portions of $J_{3}$, $J_{4}$, $J_{5}$, and $J_{12}$ that connect at least one {\it charge} site, vanishing. For $\delta = 0.25$, the $J_{1}$ and the remainder of $J_{5}$ and $J_{12}$ largely retain their original AFM characteristics, albeit with reduced strength. The $J_{4}$ changes from FM to AFM, while $J_{2}$ remains FM. Consequently, the magnetic frustration is enhanced. The magnetic ground state with $\delta = 0.25$ is expected to deviate from the DSS, with a lowered $T_{SDW}$ due to the dilution of the leading exchange interactions. Notably, the dominant $J_{3}$ is found to be sensitive to the corresponding Ni-O-Ni bond lengths and angles. The magnitude of $J_{3}$ increases with larger bond angles, supporting that $J_{3}$ are superexchange interactions facilitated by the inner-apical oxygen atoms through the Ni($d_{z^2}$)-O($p_z$)-Ni($d_{z^2}$) pathways (see Tables SIV in the Supplementary Note 4).  The $d_{z^2}$ orbitals are expected to form a singlet state, leading to the strong antiferromagnetic $J_{3}$.

When $\delta$ reaches 0.5, half of the Ni spins are converted into {\it charges}. With an ordered vacancy distribution as shown in Supplementary Note 4 Fig. S4(c), the resulting magnetic configuration effectively becomes an ideal SCS, as shown in Fig.~\ref{fig:tu2}(b). Consequently, all the nearest neighbor (NN) intralayer $J$'s ($J_{1}$ and $J_{2}$) are significantly weakened, as each $J$ connects one spin and one {\it charge} (see Supplementary Note 4 Tables SIV). This scenario naturally reproduces the theoretical model proposed in Ref.~\cite{chen2024electronic}, where for the DSS phase, all the NN intralayer interactions are artificially set to zero to obtain magnetic excitations consistent with experimental data. With $\delta = 0.5$, the corresponding $T_{SDW}$ is expected to be further lowered due to the increased presence of {\it charges}.

\begin{figure*}
	\includegraphics[scale=0.58]{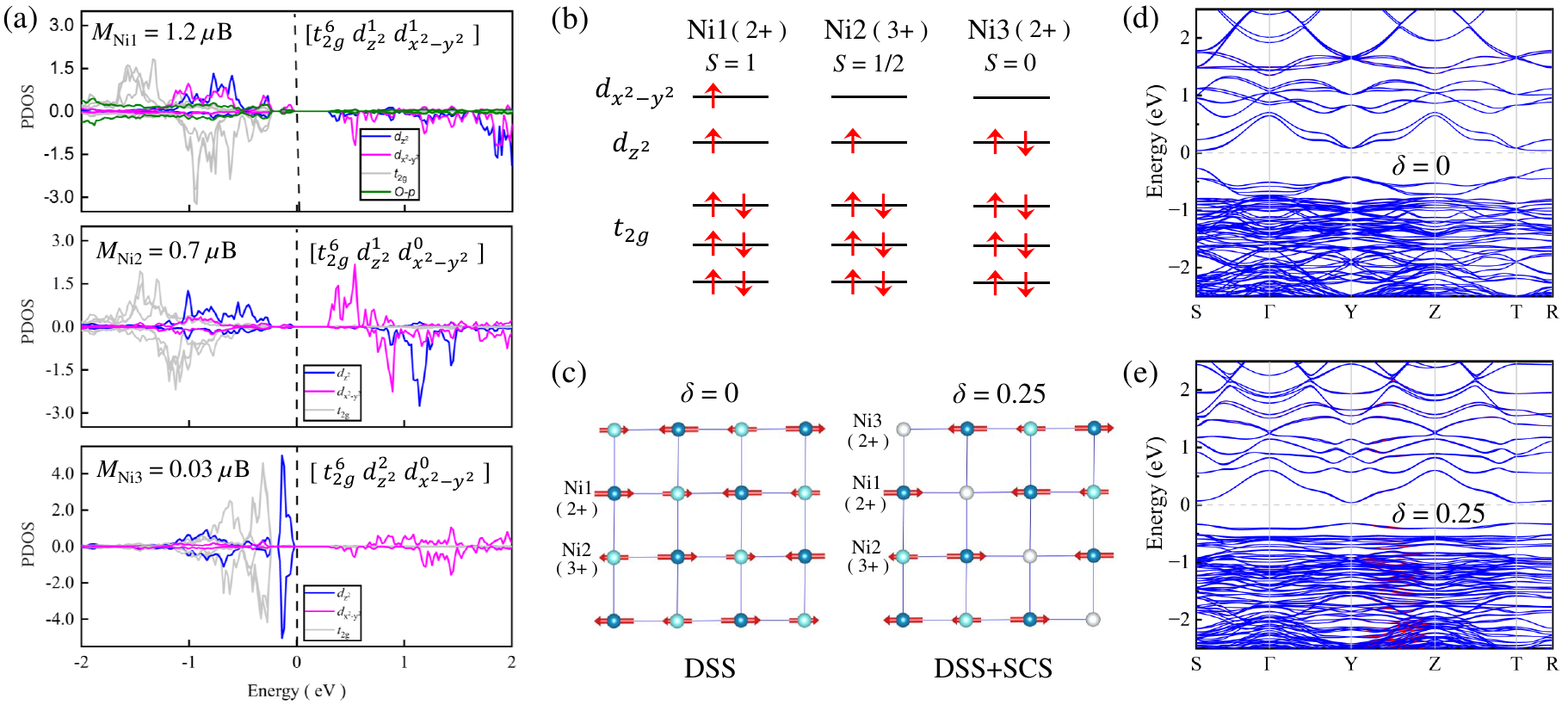}\caption{\label{fig:tu1.5} \textbf{Electronic structures of La$_3$Ni$_2$O$_{7-\delta}$}. (a) Calculated partial DOS of Ni1, Ni2 and Ni3 for $\delta = 0.25$.(b) Sketches of electron configurations of Ni1, Ni2 and Ni3. (c) Illustration of the calculated DSS phase with $\delta = 0$, showing the alternating magnetic moments of Ni1 and Ni2 and the concomitant charge order. (d) Illustration of a calculated double spin/charge stripe phase with $\delta = 0.25$, showing the concomitant charge order with Ni1, Ni2 and Ni3. Red arrows represent spins. Calculated electron band structures of (d) DSS phase with $\delta = 0$ and (e) double spin/charge stripe phase with $\delta = 0.25$. The red and blue lines represent spin-up and spin-down states, respectively.}
\end{figure*}


To confirm the impact of oxygen vacancies on the magnetic phase diagrams, we first perform MC simulations for both $\delta = 0.25$ and $\delta = 0.5$, using the $J$'s from direct DFT calculations with the ordered vacancy distributions (see Supplementary Note 4 Tables SIII and SV). Consistent with our qualitative analysis, we find that for $\delta = 0.25$, the magnetic ground state becomes noncollinear and slightly deviates from the DSS phase (see Supplementary Note 5 Fig. S6(b)). Fourier analysis shows that the $Q$ vector (0.25, 0.25) is preserved, but the $T_{SDW}$ decreases to 155 K. For $\delta = 0.5$, the simulated magnetic ground state indeed effectively becomes an ordered SCS, but with noncollinear spins and a drastically reduced $T_{SDW}$ of 48 K (see Supplementary Note 5 Fig. S7). For $\delta > 0$, the dilution of the strong interlayer interaction $J_3$ raises the transition temperature of magnetic ordering.

However, oxygen vacancies are expected to be disordered in experimental samples. Therefore, to simulate a more realistic scenario, we also perform MC simulations with randomly distributed oxygen vacancies. In this case, approximations are made to simplify the calculations, with the $J$'s directly involving the {\it charge} sites set to zero and the other $J$'s assigned values from Table~\ref{tab1}. The obtained phase diagram and magnetic structures are shown in Fig.~\ref{fig:tu3} and Supplementary Note 6 Fig. S8. As expected, the simulated $T_{SDW}$ decreases with increasing $\delta$. For $0 < \delta < 0.5$, the resulting magnetic ground state can be considered a mixture of the DSS and the SCS, characterized by a dominant global modulation vector (0.25, 0.25) (see Supplementary Note 6 Fig. S8), which is distinct from 214 nickelates, where the global wave vector continuously varies with doping~\cite{PhysRevMaterials.7.024412,PhysRevB.102.165130}. A sudden drop in $T_{SDW}$ at $\delta \sim 0.3$ marks the separation of two distinct phases. For $0< \delta < 0.3$, the magnetic ground state is dominated by the DSS with collinear spin structures (red area in Fig.~\ref{fig:tu3}), while for larger $\delta$ ($0.3 \leq \delta < 0.5$), the SCS outweighs the DSS, and the spin structures become slightly noncollinear (green area in Fig.~\ref{fig:tu3}). For $\delta \geq 0.5$, the calculated specific heat from the MC simulations shows a bump instead of a diverging peak, suggesting the presence of only short-range order. The corresponding spin structures become strongly noncollinear with spin-glass-like characteristics (blue area in Fig.~\ref{fig:tu3}).
Our simulated phase diagram provides a natural explanation to reconcile the seemingly contradictory experimental findings that both the DSS and the SCS are proposed as candidates for the SDW phase. Nevertheless, our calculations indicate that oxygen vacancies notably affect the magnetic properties of La$_3$Ni$_2$O$_7$, lowering $T_{SDW}$ and altering the ground state magnetic structures.

~\\
\noindent {\bf\large Charge order and orbital order}\\

Magnetic ordering with notable spiltting of magnetic momoments may naturally lead to concomitant charge ordering and orbital ordering. In the DSS phase, the emergence of the splitted magnetic moments of 1.2 $\mu_B$ (Ni1) and 0.7 $\mu_B$ (Ni2) can be attributed to the formation of different valence states of Ni ions. For $\delta > 0$, the presence of inner-apical oxygen vacancies introduces non-magnetic Ni ions (Ni3) in addition to Ni1 and Ni2.
Our calculated partial density of states (PDOS) of Ni1 and Ni2 with \(\delta = 0.25\) are very close to those with  \(\delta = 0\) (see Supplementary Note 7 for the PDOS with \(\delta = 0\) and Fig.~\ref{fig:tu1.5}(a) for \(\delta = 0.25\)), which suggests that the presence of moderate oxygen vacancies only has mild impact on the electronic configurations of Ni1 and Ni2. Therefore, for simplicity, we mainly discuss the PDOS with \(\delta = 0.25\) , in which case, Ni1, Ni2 and Ni3 are all present.

 Our DFT calculated PDOS indicates that the Ni1 has a valence state of 2+ with an electron configuration of $(t_{2g}^6 \, d_{z^2}^1 \, d_{x^2-y^2}^1)$, while the Ni2 has an electron configuration of $(t_{2g}^6 \, d_{z^2}^1)$, corresponding to Ni$^{3+}$(see Fig.~\ref{fig:tu1.5}(a) and (b)).  In the perfect DSS phase with $\delta = 0$, the orbital ordering originates from the splitting of the $d_{z^2}$ and $d_{x^2-y^2}$ orbitals due to the alternating compression and elongation of NiO$_6$ octahedra, which is driven by spin-lattice coupling under the DSS magnetic order~\cite{labollita2024assessing,zhang2024emergent}.  With moderate oxygen deficiency, although the NiO$_6$ octahedra containing Ni3 are disrupted, the electron occupations of Ni3 can still be approximated using the $t_{2g}$ and $e_g$ levels, assuming the original crystal field. Ni3 is found to have a valence state of 2+ with a low-spin electron configuration of $(t_{2g}^6 \, d_{z^2}^2)$.
Therefore, the loss of magnetic moment of Ni3 is due to the full occupation of the $d_{z^2}$ orbitals. Intuitively, the energies of Ni3 $d_{z^2}$ orbitals should be the most affected by the absence of inner-apical oxygens, since these $d_{z^2}$ orbitals point towards the vacancies. In this case, the $d_{z^2}$ levels of Ni3 are lowered to the extent that the energy drop surpasses the spin-pairing energy. On the other hand, the inner-apical oxygen vacancy disrupts the interlayer super-exchange interaction. Although the in-plane interactions remain largely intact, they are rather weak compared to the interlayer interaction. As a result, the Ni spins nearest to the vacancies can be approximated as nearly free spins, which may provide another perspective to interpret the disappearance of magnetic moment at Ni3 sites.
Consequently, the magnetic ordering in the phase diagram shown in Fig.~\ref{fig:tu2} naturally leads to concurrent charge and orbital ordering, as illustrated in Fig.~\ref{fig:tu1.5}(c).


The electronic structures for the DSS phase with $\delta = 0$ and the double spin/charge stripe phase with $\delta = 0.25$ are shown in Fig.~\ref{fig:tu1.5}(d) and Fig.~\ref{fig:tu1.5}(e), respectively. The two phases are both semiconducting with a small gap of 0.31 eV and 0.36 eV respectively, which is consistent with some experimental observations~\cite{hou2023emergence,zhang2024high} and recent theoretical studies~\cite{labollita2024assessing,zhang2024emergent}. Especially, a gap like feature in the electronic struture of La$_3$Ni$_2$O$_{7}$ is observed via scanning tunneling microscopy/spectroscopy~\cite{,PhysRevB.110.134520}. In contrast, the DFT gap vanishes in the nonmagnetic ( not spin polarized ) phase, even with larger \(U\) values, suggesting that the gap is opened by the onset of spin density wave, together with the accompanying charge order and orbital order. The weak spin spliting in the case of $\delta = 0.25$ is due to that the Ni3 still has a weak magnetic moment of 0.03 $\mu_B$. 
It is noteworthy that some of the Fe-based superconductors, such as FeTe~\cite{dai2015antiferromagnetic} and Fe$_{1+y}$Te$_{1-x}$Se$_{x}$~\cite{fang2009theory}, also have a magnetic ground state characterized by a DSS configuration, yet their electronic structure exhibits metallic properties~\cite{subedi2008density}.

\begin{figure}
	\includegraphics[scale=0.10]{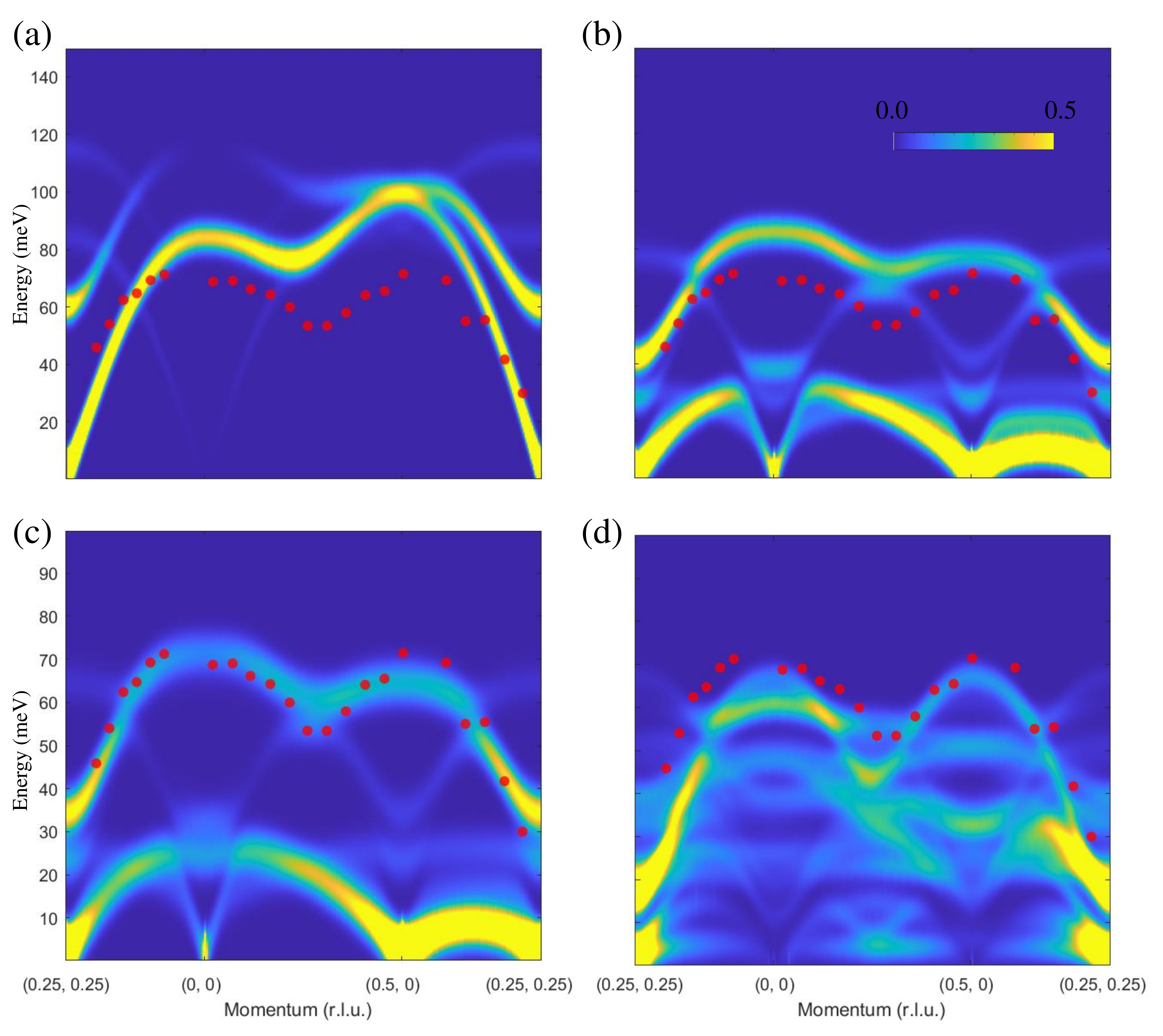}\caption{\label{fig:tu4}\textbf{Magnetic excitation spectra of La$_3$Ni$_2$O$_{7-\delta}$}. The color maps represent calulated spin wave of La$_3$Ni$_2$O$_{7-\delta}$ with (a) $\delta = 0$ (b) $\delta = 0.25$ with an ordered oxygen vacancy distribution.  (c) $\delta = 0.25$ with an ordered oxygen vacancy distribution and all $J$'s scaled by 0.83. (d) $\delta = 0.25$ with a random distribution of splitted $J$'s. Red dots represent experimental RIXS data in Ref.~\cite{zhou2024electronic}.}
\end{figure}

~\\
\noindent {\bf\large Magnetic excitations}\\
We calculate the magnetic excitation spectra of La$_3$Ni$_2$O$_{7}$ using linear spin wave theory as implemented in the SpinW software package~\cite{toth2015linear}, considering all the calculated exchange interactions \(J\)'s. For $\delta = 0$, the calculated spectra along (0.25, 0.25)-(0, 0)-(0.5, 0)-(0.25, 0.25) exhibit characteristics consistent with experimental RIXS data in Ref.~\cite{zhou2024electronic}, but with the band top overestimated by approximately a factor of two, as shown in Fig.~\ref{fig:tu4}(a). This overestimation aligns with the similarly overestimated $T_{SDW}$ of 232 K. 
With $\delta = 0.25$, the band top lowers to about 90 meV (shown in Fig.~\ref{fig:tu4}(b)), bringing it closer to the experimental data due to the dilution and weakening of the leading $J$ values. If we further manually reduce all the $J$ values by 17\%, the resulting band top falls almost on top of the experimental dispersion (see Fig.~\ref{fig:tu4}(c)). However, the experimental spectra around (0.5, 0) are nearly symmetric with those around (0, 0), while the theoretical spectra around (0.5, 0) are notably lower. To further consider the effect of disordered oxygen vacancies, we calculate the spin wave spectra with a random splitting of $J$ values using a 2 $\times$ 2 $\times$ 1 supercell, corresponding to a random distribution of vacancies. As shown in Fig.~\ref{fig:tu4}(d), introducing randomness not only effectively lowers the band top of the spin wave dispersions but also makes the band top more symmetric, resulting in better agreement with the experimental data.

~\\
\noindent{\bf\large{Discussions}}\\

It is worth emphasizing that, given the complexities of dealing with correlated electrons, our study may only provides a semiquantitative analysis of the magnetic properties of La$_3$Ni$_2$O$_{7-\delta}$. The calculated magnetic moment appears larger than the experimental observed ones, with even larger values with increased \(U\). However, based on our calculations, the presence of oxygen vacancies may introduce magnetic noncollinearity and spinless charge sites, which can lower the overall magnetic moments, bringing our results closer to the experimental observations. In this context, although La\(_3\)Ni\(_2\)O\(_{7-\delta}\) may also exhibit zero-point quantum fluctuations similar to those observed in 214 nickelates~\cite{PhysRevLett.127.275301,PhysRevB.105.195147}, which could also help to reduce both the magnetic moment and the spin wave dispersion energy spread, we believe that our DFT + \(U\) calculations, combined with classical MC simulations still provide a reasonable interpretation for the magnetic properties of La\(_3\)Ni\(_2\)O\(_{7-\delta}\).

Our results indicate that the $T_{SDW}$ and spin wave excitation energies are sensitive to the concentration of oxygen vacancies. At first glance, these conclusions appear to contradict experimental findings, where $T_{SDW}$ is consistently measured around 150 K across various samples. However, as shown in Refs.~\cite{dan2024spin, dong2023visualization}, the distribution of oxygen vacancies in experimental samples is highly inhomogeneous. For instance, $\delta$ varies from approximately 0.04 to 0.32 in different regions of a single sample used in Ref.~\cite{dong2023visualization}. 
A plausible speculation is that the bulk volume of different samples has a similar moderate $\delta$ value, e.g., $\delta \sim 0.20$ as observed in Ref.~\cite{dong2023visualization}, which supports a quasi-long-range magnetic ordering with a wave vector of (0.25, 0.25), resulting in a consistent $T_{SDW}$. The discrepancies lie in the vacancy distribution in other regions. High-$\delta$ regions can induce spin-glass-like short-range orders at low temperatures, preventing the formation of long-range orders across the entire sample. Conversely, regions with smaller $\delta$ can exhibit magnetic ordering up to higher temperatures. This picture aligns with NMR observations, where critical spin fluctuations in regions away from oxygen vacancies persist up to 200 K, while below 50 K, glassy spin dynamics are observed~\cite{dan2024spin}.
%
 
In conclusion, our calculations reveal strong nearest neighbor interlayer interactions, with amplitudes one order of magnitude larger than the other interactions. For $\delta=0$, our calculations favor the DSS phase with a (0.25, 0.25) in-plane modulation vector as the magnetic ground state. Oxygen vacancies effectively turn the Ni magnetic moments in the vicinity into {\it charge} sites. Therefore, with moderate $\delta$ values, our theoretical SDW phase is found to possess characteristics of both the DSS and the SCS, maintaining (0.25, 0.25) as the dominant modulation vector. This picture may provide a natural explanation to reconcile the seemingly contradictory experimental findings that both the DSS and the SCS are proposed as candidates for the SDW phase. 
At high concentrations of oxygen vacancies, such as $\delta=0.5$, we anticipate a short-range ordered ground state with spin-glass-like magnetic structures, leading to the destruction of the (0.25, 0.25) modulation vector. The presence of oxygen vacancies also lowers the $T_{SDW}$ due to the dilution of dominant exchange interactions. Notably, the magnetic ordering induces concurrent charge and orbital ordering, facilitated by spin-lattice coupling under the low-symmetry magnetic phase. Moreover, we find that the effect of a random distribution of oxygen vacancies may need to be considered to obtain a spin wave spectrum consistent with experiments. We further provide a plausible explanation for the experimental observations that the measured $T_{SDW}$ seems not sensitive to different samples and the lack of direct evidence for long-range magnetic ordering.

{\it Note added}: We notice two related works on the SDW of La\(_3\)Ni\(_2\)O\(_{7-\delta}\)~\cite{labollita2024assessing,zhang2024emergent}, which were posted around a week prior to our first submission. The two works focus on both the SDW under ambient pressures and high pressures, but without considering oxygen vacancies, while our work focus on the SDW under ambient pressures and the impact of oxygen deficiencies on the magnetic properties.



\begin{thebibliography}{74}%
\makeatletter
\providecommand \@ifxundefined [1]{%
 \@ifx{#1\undefined}
}%
\providecommand \@ifnum [1]{%
 \ifnum #1\expandafter \@firstoftwo
 \else \expandafter \@secondoftwo
 \fi
}%
\providecommand \@ifx [1]{%
 \ifx #1\expandafter \@firstoftwo
 \else \expandafter \@secondoftwo
 \fi
}%
\providecommand \natexlab [1]{#1}%
\providecommand \enquote  [1]{``#1''}%
\providecommand \bibnamefont  [1]{#1}%
\providecommand \bibfnamefont [1]{#1}%
\providecommand \citenamefont [1]{#1}%
\providecommand \href@noop [0]{\@secondoftwo}%
\providecommand \href [0]{\begingroup \@sanitize@url \@href}%
\providecommand \@href[1]{\@@startlink{#1}\@@href}%
\providecommand \@@href[1]{\endgroup#1\@@endlink}%
\providecommand \@sanitize@url [0]{\catcode `\\12\catcode `\$12\catcode
  `\&12\catcode `\#12\catcode `\^12\catcode `\_12\catcode `\%12\relax}%
\providecommand \@@startlink[1]{}%
\providecommand \@@endlink[0]{}%
\providecommand \url  [0]{\begingroup\@sanitize@url \@url }%
\providecommand \@url [1]{\endgroup\@href {#1}{\urlprefix }}%
\providecommand \urlprefix  [0]{URL }%
\providecommand \Eprint [0]{\href }%
\providecommand \doibase [0]{http://dx.doi.org/}%
\providecommand \selectlanguage [0]{\@gobble}%
\providecommand \bibinfo  [0]{\@secondoftwo}%
\providecommand \bibfield  [0]{\@secondoftwo}%
\providecommand \translation [1]{[#1]}%
\providecommand \BibitemOpen [0]{}%
\providecommand \bibitemStop [0]{}%
\providecommand \bibitemNoStop [0]{.\EOS\space}%
\providecommand \EOS [0]{\spacefactor3000\relax}%
\providecommand \BibitemShut  [1]{\csname bibitem#1\endcsname}%
\let\auto@bib@innerbib\@empty
\bibitem [{\citenamefont {Sun}\ \emph {et~al.}(2023)\citenamefont {Sun},
  \citenamefont {Huo}, \citenamefont {Hu}, \citenamefont {Li}, \citenamefont
  {Liu}, \citenamefont {Han}, \citenamefont {Tang}, \citenamefont {Mao},
  \citenamefont {Yang}, \citenamefont {Wang} \emph {et~al.}}]{20}%
  \BibitemOpen
  \bibfield  {author} {\bibinfo {author} {\bibfnamefont {Hualei}\ \bibnamefont
  {Sun}}, \bibinfo {author} {\bibfnamefont {Mengwu}\ \bibnamefont {Huo}},
  \bibinfo {author} {\bibfnamefont {Xunwu}\ \bibnamefont {Hu}}, \bibinfo
  {author} {\bibfnamefont {Jingyuan}\ \bibnamefont {Li}}, \bibinfo {author}
  {\bibfnamefont {Zengjia}\ \bibnamefont {Liu}}, \bibinfo {author}
  {\bibfnamefont {Yifeng}\ \bibnamefont {Han}}, \bibinfo {author}
  {\bibfnamefont {Lingyun}\ \bibnamefont {Tang}}, \bibinfo {author}
  {\bibfnamefont {Zhongquan}\ \bibnamefont {Mao}}, \bibinfo {author}
  {\bibfnamefont {Pengtao}\ \bibnamefont {Yang}}, \bibinfo {author}
  {\bibfnamefont {Bosen}\ \bibnamefont {Wang}},  \emph {et~al.},\ }\bibfield
  {title} {\enquote {\bibinfo {title} {{Signatures of superconductivity near 80
  K in a nickelate under high pressure}},}\ }\href {\doibase
  10.1038/s41586-023-06408-7} {\bibfield  {journal} {\bibinfo  {journal}
  {Nature}\ ,\ \bibinfo {pages} {1--3}} (\bibinfo {year} {2023})}\BibitemShut
  {NoStop}%
\bibitem [{\citenamefont {Hou}\ \emph {et~al.}(2023{\natexlab{a}})\citenamefont
  {Hou}, \citenamefont {Yang}, \citenamefont {Liu}, \citenamefont {Li},
  \citenamefont {Shan}, \citenamefont {Ma}, \citenamefont {Wang}, \citenamefont
  {Wang}, \citenamefont {Guo}, \citenamefont {Sun} \emph {et~al.}}]{21}%
  \BibitemOpen
  \bibfield  {author} {\bibinfo {author} {\bibfnamefont {J}~\bibnamefont
  {Hou}}, \bibinfo {author} {\bibfnamefont {PT}~\bibnamefont {Yang}}, \bibinfo
  {author} {\bibfnamefont {ZY}~\bibnamefont {Liu}}, \bibinfo {author}
  {\bibfnamefont {JY}~\bibnamefont {Li}}, \bibinfo {author} {\bibfnamefont
  {PF}~\bibnamefont {Shan}}, \bibinfo {author} {\bibfnamefont {L}~\bibnamefont
  {Ma}}, \bibinfo {author} {\bibfnamefont {G}~\bibnamefont {Wang}}, \bibinfo
  {author} {\bibfnamefont {NN}~\bibnamefont {Wang}}, \bibinfo {author}
  {\bibfnamefont {HZ}~\bibnamefont {Guo}}, \bibinfo {author} {\bibfnamefont
  {JP}~\bibnamefont {Sun}},  \emph {et~al.},\ }\bibfield  {title} {\enquote
  {\bibinfo {title} {{Emergence of high-temperature superconducting phase in
  the pressurized La$_{3}$Ni$_{2}$O$_{7}$ crystals}},}\ }\href {\doibase
  10.48550/arXiv.2307.09865} {\bibfield  {journal} {\bibinfo  {journal} {arXiv
  preprint arXiv:2307.09865}\ } (\bibinfo {year} {2023}{\natexlab{a}}),\
  10.48550/arXiv.2307.09865}\BibitemShut {NoStop}%
\bibitem [{\citenamefont {Zhang}\ \emph
  {et~al.}(2024{\natexlab{a}})\citenamefont {Zhang}, \citenamefont {Su},
  \citenamefont {Huang}, \citenamefont {Shan}, \citenamefont {Sun},
  \citenamefont {Huo}, \citenamefont {Ye}, \citenamefont {Zhang}, \citenamefont
  {Yang}, \citenamefont {Xu} \emph {et~al.}}]{zhang2024high}%
  \BibitemOpen
  \bibfield  {author} {\bibinfo {author} {\bibfnamefont {Yanan}\ \bibnamefont
  {Zhang}}, \bibinfo {author} {\bibfnamefont {Dajun}\ \bibnamefont {Su}},
  \bibinfo {author} {\bibfnamefont {Yanen}\ \bibnamefont {Huang}}, \bibinfo
  {author} {\bibfnamefont {Zhaoyang}\ \bibnamefont {Shan}}, \bibinfo {author}
  {\bibfnamefont {Hualei}\ \bibnamefont {Sun}}, \bibinfo {author}
  {\bibfnamefont {Mengwu}\ \bibnamefont {Huo}}, \bibinfo {author}
  {\bibfnamefont {Kaixin}\ \bibnamefont {Ye}}, \bibinfo {author} {\bibfnamefont
  {Jiawen}\ \bibnamefont {Zhang}}, \bibinfo {author} {\bibfnamefont {Zihan}\
  \bibnamefont {Yang}}, \bibinfo {author} {\bibfnamefont {Yongkang}\
  \bibnamefont {Xu}},  \emph {et~al.},\ }\bibfield  {title} {\enquote {\bibinfo
  {title} {{High-temperature superconductivity with zero resistance and
  strange-metal behaviour in La$_3$Ni$_2$O$_{7-\delta}$}},}\ }\href {\doibase
  10.1038/s41567-024-02515-y} {\bibfield  {journal} {\bibinfo  {journal}
  {Nature Physics}\ ,\ \bibinfo {pages} {1--5}} (\bibinfo {year}
  {2024}{\natexlab{a}})}\BibitemShut {NoStop}%
\bibitem [{\citenamefont {Wang}\ \emph {et~al.}(2024)\citenamefont {Wang},
  \citenamefont {Wang}, \citenamefont {Shen}, \citenamefont {Hou},
  \citenamefont {Ma}, \citenamefont {Shi}, \citenamefont {Ren}, \citenamefont
  {Gu}, \citenamefont {Ma}, \citenamefont {Yang} \emph
  {et~al.}}]{wang2024pressure}%
  \BibitemOpen
  \bibfield  {author} {\bibinfo {author} {\bibfnamefont {Gang}\ \bibnamefont
  {Wang}}, \bibinfo {author} {\bibfnamefont {NN}~\bibnamefont {Wang}}, \bibinfo
  {author} {\bibfnamefont {XL}~\bibnamefont {Shen}}, \bibinfo {author}
  {\bibfnamefont {J}~\bibnamefont {Hou}}, \bibinfo {author} {\bibfnamefont
  {L}~\bibnamefont {Ma}}, \bibinfo {author} {\bibfnamefont {LF}~\bibnamefont
  {Shi}}, \bibinfo {author} {\bibfnamefont {ZA}~\bibnamefont {Ren}}, \bibinfo
  {author} {\bibfnamefont {YD}~\bibnamefont {Gu}}, \bibinfo {author}
  {\bibfnamefont {HM}~\bibnamefont {Ma}}, \bibinfo {author} {\bibfnamefont
  {PT}~\bibnamefont {Yang}},  \emph {et~al.},\ }\bibfield  {title} {\enquote
  {\bibinfo {title} {{Pressure-Induced Superconductivity In Polycrystalline
  La$_3$Ni$_2$O$_{7-\delta}$}},}\ }\href {\doibase 10.1103/PhysRevX.14.011040}
  {\bibfield  {journal} {\bibinfo  {journal} {Physical Review X}\ }\textbf
  {\bibinfo {volume} {14}},\ \bibinfo {pages} {011040} (\bibinfo {year}
  {2024})}\BibitemShut {NoStop}%
\bibitem [{\citenamefont {Zhang}\ \emph
  {et~al.}(2023{\natexlab{a}})\citenamefont {Zhang}, \citenamefont {Pei},
  \citenamefont {Wang}, \citenamefont {Zhao}, \citenamefont {Li}, \citenamefont
  {Cao}, \citenamefont {Zhu}, \citenamefont {Wu},\ and\ \citenamefont
  {Qi}}]{zhang2023effects}%
  \BibitemOpen
  \bibfield  {author} {\bibinfo {author} {\bibfnamefont {Mingxin}\ \bibnamefont
  {Zhang}}, \bibinfo {author} {\bibfnamefont {Cuiying}\ \bibnamefont {Pei}},
  \bibinfo {author} {\bibfnamefont {Qi}~\bibnamefont {Wang}}, \bibinfo {author}
  {\bibfnamefont {Yi}~\bibnamefont {Zhao}}, \bibinfo {author} {\bibfnamefont
  {Changhua}\ \bibnamefont {Li}}, \bibinfo {author} {\bibfnamefont {Weizheng}\
  \bibnamefont {Cao}}, \bibinfo {author} {\bibfnamefont {Shihao}\ \bibnamefont
  {Zhu}}, \bibinfo {author} {\bibfnamefont {Juefei}\ \bibnamefont {Wu}}, \ and\
  \bibinfo {author} {\bibfnamefont {Yanpeng}\ \bibnamefont {Qi}},\ }\bibfield
  {title} {\enquote {\bibinfo {title} {Effects of pressure and doping on
  ruddlesden-popper phases},}\ }\href {\doibase 10.1016/j.jmst.2023.11.011}
  {\bibfield  {journal} {\bibinfo  {journal} {arXiv preprint arXiv:2309.01651}\
  } (\bibinfo {year} {2023}{\natexlab{a}}),\
  10.1016/j.jmst.2023.11.011}\BibitemShut {NoStop}%
\bibitem [{\citenamefont {Zhou}\ \emph {et~al.}(2023)\citenamefont {Zhou},
  \citenamefont {Guo}, \citenamefont {Cai}, \citenamefont {Sun}, \citenamefont
  {Wang}, \citenamefont {Zhao}, \citenamefont {Han}, \citenamefont {Chen},
  \citenamefont {Wu}, \citenamefont {Ding} \emph {et~al.}}]{zhou2023evidence}%
  \BibitemOpen
  \bibfield  {author} {\bibinfo {author} {\bibfnamefont {Yazhou}\ \bibnamefont
  {Zhou}}, \bibinfo {author} {\bibfnamefont {Jing}\ \bibnamefont {Guo}},
  \bibinfo {author} {\bibfnamefont {Shu}\ \bibnamefont {Cai}}, \bibinfo
  {author} {\bibfnamefont {Hualei}\ \bibnamefont {Sun}}, \bibinfo {author}
  {\bibfnamefont {Pengyu}\ \bibnamefont {Wang}}, \bibinfo {author}
  {\bibfnamefont {Jinyu}\ \bibnamefont {Zhao}}, \bibinfo {author}
  {\bibfnamefont {Jinyu}\ \bibnamefont {Han}}, \bibinfo {author} {\bibfnamefont
  {Xintian}\ \bibnamefont {Chen}}, \bibinfo {author} {\bibfnamefont
  {Qi}~\bibnamefont {Wu}}, \bibinfo {author} {\bibfnamefont {Yang}\
  \bibnamefont {Ding}},  \emph {et~al.},\ }\bibfield  {title} {\enquote
  {\bibinfo {title} {{Evidence of filamentary superconductivity in pressurized
  La$_3$Ni$_2$O$_{7}$ single crystals}},}\ }\href {\doibase
  10.48550/arXiv.2311.12361} {\bibfield  {journal} {\bibinfo  {journal} {arXiv
  preprint arXiv:2311.12361}\ } (\bibinfo {year} {2023}),\
  10.48550/arXiv.2311.12361}\BibitemShut {NoStop}%
\bibitem [{\citenamefont {Luo}\ \emph {et~al.}(2023)\citenamefont {Luo},
  \citenamefont {Hu}, \citenamefont {Wang}, \citenamefont {W\'u},\ and\
  \citenamefont {Yao}}]{luo2023bilayer}%
  \BibitemOpen
  \bibfield  {author} {\bibinfo {author} {\bibfnamefont {Zhihui}\ \bibnamefont
  {Luo}}, \bibinfo {author} {\bibfnamefont {Xunwu}\ \bibnamefont {Hu}},
  \bibinfo {author} {\bibfnamefont {Meng}\ \bibnamefont {Wang}}, \bibinfo
  {author} {\bibfnamefont {W\'ei}\ \bibnamefont {W\'u}}, \ and\ \bibinfo
  {author} {\bibfnamefont {Dao-Xin}\ \bibnamefont {Yao}},\ }\bibfield  {title}
  {\enquote {\bibinfo {title} {{Bilayer two-orbital model of
  La$_{3}$Ni$_{2}$O$_{7}$ under pressure}},}\ }\href {\doibase
  10.1103/PhysRevLett.131.126001} {\bibfield  {journal} {\bibinfo  {journal}
  {Phys. Rev. Lett.}\ }\textbf {\bibinfo {volume} {131}},\ \bibinfo {pages}
  {126001} (\bibinfo {year} {2023})}\BibitemShut {NoStop}%
\bibitem [{\citenamefont {Yang}\ \emph {et~al.}(2024)\citenamefont {Yang},
  \citenamefont {Sun}, \citenamefont {Hu}, \citenamefont {Xie}, \citenamefont
  {Miao}, \citenamefont {Luo}, \citenamefont {Chen}, \citenamefont {Liang},
  \citenamefont {Zhu}, \citenamefont {Qu} \emph {et~al.}}]{yang2024orbital}%
  \BibitemOpen
  \bibfield  {author} {\bibinfo {author} {\bibfnamefont {Jiangang}\
  \bibnamefont {Yang}}, \bibinfo {author} {\bibfnamefont {Hualei}\ \bibnamefont
  {Sun}}, \bibinfo {author} {\bibfnamefont {Xunwu}\ \bibnamefont {Hu}},
  \bibinfo {author} {\bibfnamefont {Yuyang}\ \bibnamefont {Xie}}, \bibinfo
  {author} {\bibfnamefont {Taimin}\ \bibnamefont {Miao}}, \bibinfo {author}
  {\bibfnamefont {Hailan}\ \bibnamefont {Luo}}, \bibinfo {author}
  {\bibfnamefont {Hao}\ \bibnamefont {Chen}}, \bibinfo {author} {\bibfnamefont
  {Bo}~\bibnamefont {Liang}}, \bibinfo {author} {\bibfnamefont {Wenpei}\
  \bibnamefont {Zhu}}, \bibinfo {author} {\bibfnamefont {Gexing}\ \bibnamefont
  {Qu}},  \emph {et~al.},\ }\bibfield  {title} {\enquote {\bibinfo {title}
  {{Orbital-dependent electron correlation in double-layer nickelate
  La$_{3}$Ni$_{2}$O$_{7}$}},}\ }\href {\doibase 10.1038/s41467-024-48701-7}
  {\bibfield  {journal} {\bibinfo  {journal} {Nature Communications}\ }\textbf
  {\bibinfo {volume} {15}},\ \bibinfo {pages} {4373} (\bibinfo {year}
  {2024})}\BibitemShut {NoStop}%
\bibitem [{\citenamefont {Luo}\ \emph {et~al.}(2024)\citenamefont {Luo},
  \citenamefont {Lv}, \citenamefont {Wang}, \citenamefont {W{\'u}},\ and\
  \citenamefont {Yao}}]{luo2024high}%
  \BibitemOpen
  \bibfield  {author} {\bibinfo {author} {\bibfnamefont {Zhihui}\ \bibnamefont
  {Luo}}, \bibinfo {author} {\bibfnamefont {Biao}\ \bibnamefont {Lv}}, \bibinfo
  {author} {\bibfnamefont {Meng}\ \bibnamefont {Wang}}, \bibinfo {author}
  {\bibfnamefont {W{\'e}i}\ \bibnamefont {W{\'u}}}, \ and\ \bibinfo {author}
  {\bibfnamefont {Dao-Xin}\ \bibnamefont {Yao}},\ }\bibfield  {title} {\enquote
  {\bibinfo {title} {{High-T$_c$ superconductivity in La$_{3}$Ni$_{2}$O$_{7}$
  based on the bilayer two-orbital t-J model}},}\ }\href {\doibase
  10.1038/s41535-024-00668-w} {\bibfield  {journal} {\bibinfo  {journal} {npj
  Quantum Materials}\ }\textbf {\bibinfo {volume} {9}},\ \bibinfo {pages} {61}
  (\bibinfo {year} {2024})}\BibitemShut {NoStop}%
\bibitem [{\citenamefont {Kang}\ \emph {et~al.}(2023)\citenamefont {Kang},
  \citenamefont {Melnick}, \citenamefont {Semon}, \citenamefont {Ryee},
  \citenamefont {Han}, \citenamefont {Kotliar},\ and\ \citenamefont
  {Choi}}]{kang2023infinite}%
  \BibitemOpen
  \bibfield  {author} {\bibinfo {author} {\bibfnamefont {Byungkyun}\
  \bibnamefont {Kang}}, \bibinfo {author} {\bibfnamefont {Corey}\ \bibnamefont
  {Melnick}}, \bibinfo {author} {\bibfnamefont {Patrick}\ \bibnamefont
  {Semon}}, \bibinfo {author} {\bibfnamefont {Siheon}\ \bibnamefont {Ryee}},
  \bibinfo {author} {\bibfnamefont {Myung~Joon}\ \bibnamefont {Han}}, \bibinfo
  {author} {\bibfnamefont {Gabriel}\ \bibnamefont {Kotliar}}, \ and\ \bibinfo
  {author} {\bibfnamefont {Sangkook}\ \bibnamefont {Choi}},\ }\bibfield
  {title} {\enquote {\bibinfo {title} {{Infinite-layer nickelates as Ni-e$_g$
  Hund’s metals}},}\ }\href {\doibase 10.1038/s41535-023-00568-5} {\bibfield
  {journal} {\bibinfo  {journal} {npj Quantum Materials}\ }\textbf {\bibinfo
  {volume} {8}},\ \bibinfo {pages} {35} (\bibinfo {year} {2023})}\BibitemShut
  {NoStop}%
\bibitem [{\citenamefont {Christiansson}\ \emph {et~al.}(2023)\citenamefont
  {Christiansson}, \citenamefont {Petocchi},\ and\ \citenamefont
  {Werner}}]{PhysRevLett.131.206501}%
  \BibitemOpen
  \bibfield  {author} {\bibinfo {author} {\bibfnamefont {Viktor}\ \bibnamefont
  {Christiansson}}, \bibinfo {author} {\bibfnamefont {Francesco}\ \bibnamefont
  {Petocchi}}, \ and\ \bibinfo {author} {\bibfnamefont {Philipp}\ \bibnamefont
  {Werner}},\ }\bibfield  {title} {\enquote {\bibinfo {title} {{Correlated
  Electronic Structure of La$_3$Ni$_2$O$_7$ under Pressure}},}\ }\href
  {\doibase 10.1103/PhysRevLett.131.206501} {\bibfield  {journal} {\bibinfo
  {journal} {Phys. Rev. Lett.}\ }\textbf {\bibinfo {volume} {131}},\ \bibinfo
  {pages} {206501} (\bibinfo {year} {2023})}\BibitemShut {NoStop}%
\bibitem [{\citenamefont {Lechermann}\ \emph {et~al.}(2023)\citenamefont
  {Lechermann}, \citenamefont {Gondolf}, \citenamefont {B{\"o}tzel},\ and\
  \citenamefont {Eremin}}]{electronic2}%
  \BibitemOpen
  \bibfield  {author} {\bibinfo {author} {\bibfnamefont {Frank}\ \bibnamefont
  {Lechermann}}, \bibinfo {author} {\bibfnamefont {Jannik}\ \bibnamefont
  {Gondolf}}, \bibinfo {author} {\bibfnamefont {Steffen}\ \bibnamefont
  {B{\"o}tzel}}, \ and\ \bibinfo {author} {\bibfnamefont {Ilya~M}\ \bibnamefont
  {Eremin}},\ }\bibfield  {title} {\enquote {\bibinfo {title} {{Electronic
  correlations and superconducting instability in La$_3$Ni$_2$O$_{7}$ under
  high pressure}},}\ }\href {\doibase 10.48550/arXiv.2306.05121} {\bibfield
  {journal} {\bibinfo  {journal} {arXiv preprint arXiv:2306.05121}\ } (\bibinfo
  {year} {2023}),\ 10.48550/arXiv.2306.05121}\BibitemShut {NoStop}%
\bibitem [{\citenamefont {Zhang}\ \emph
  {et~al.}(2023{\natexlab{b}})\citenamefont {Zhang}, \citenamefont {Lin},
  \citenamefont {Moreo},\ and\ \citenamefont {Dagotto}}]{electronic3}%
  \BibitemOpen
  \bibfield  {author} {\bibinfo {author} {\bibfnamefont {Yang}\ \bibnamefont
  {Zhang}}, \bibinfo {author} {\bibfnamefont {Ling-Fang}\ \bibnamefont {Lin}},
  \bibinfo {author} {\bibfnamefont {Adriana}\ \bibnamefont {Moreo}}, \ and\
  \bibinfo {author} {\bibfnamefont {Elbio}\ \bibnamefont {Dagotto}},\
  }\bibfield  {title} {\enquote {\bibinfo {title} {{Electronic structure,
  orbital-selective behavior, and magnetic tendencies in the bilayer nickelate
  superconductor La$_3$Ni$_2$O$_{7}$ under pressure}},}\ }\href {\doibase
  10.48550/arXiv.2306.03231} {\bibfield  {journal} {\bibinfo  {journal} {arXiv
  preprint arXiv:2306.03231}\ } (\bibinfo {year} {2023}{\natexlab{b}}),\
  10.48550/arXiv.2306.03231}\BibitemShut {NoStop}%
\bibitem [{\citenamefont {Cao}\ and\ \citenamefont {Yang}(2023)}]{electronic4}%
  \BibitemOpen
  \bibfield  {author} {\bibinfo {author} {\bibfnamefont {Yingying}\
  \bibnamefont {Cao}}\ and\ \bibinfo {author} {\bibfnamefont {Yi-feng}\
  \bibnamefont {Yang}},\ }\bibfield  {title} {\enquote {\bibinfo {title} {{Flat
  bands promoted by Hund's rule coupling in the candidate double-layer
  high-temperature superconductor La$_3$Ni$_2$O$_{7}$}},}\ }\href {\doibase
  10.48550/arXiv.2307.06806} {\bibfield  {journal} {\bibinfo  {journal} {arXiv
  preprint arXiv:2307.06806}\ } (\bibinfo {year} {2023}),\
  10.48550/arXiv.2307.06806}\BibitemShut {NoStop}%
\bibitem [{\citenamefont {Jiang}\ \emph {et~al.}(2023)\citenamefont {Jiang},
  \citenamefont {Wang},\ and\ \citenamefont {Zhang}}]{electronic5}%
  \BibitemOpen
  \bibfield  {author} {\bibinfo {author} {\bibfnamefont {Kun}\ \bibnamefont
  {Jiang}}, \bibinfo {author} {\bibfnamefont {Ziqiang}\ \bibnamefont {Wang}}, \
  and\ \bibinfo {author} {\bibfnamefont {Fu-Chun}\ \bibnamefont {Zhang}},\
  }\bibfield  {title} {\enquote {\bibinfo {title} {{High Temperature
  Superconductivity in La$_3$Ni$_2$O$_{7}$}},}\ }\href {\doibase
  10.48550/arXiv.2308.06771} {\bibfield  {journal} {\bibinfo  {journal} {arXiv
  preprint arXiv:2308.06771}\ } (\bibinfo {year} {2023}),\
  10.48550/arXiv.2308.06771}\BibitemShut {NoStop}%
\bibitem [{\citenamefont {Huang}\ \emph {et~al.}(2023)\citenamefont {Huang},
  \citenamefont {Wang},\ and\ \citenamefont {Zhou}}]{electronic6}%
  \BibitemOpen
  \bibfield  {author} {\bibinfo {author} {\bibfnamefont {Junkang}\ \bibnamefont
  {Huang}}, \bibinfo {author} {\bibfnamefont {ZD}~\bibnamefont {Wang}}, \ and\
  \bibinfo {author} {\bibfnamefont {Tao}\ \bibnamefont {Zhou}},\ }\bibfield
  {title} {\enquote {\bibinfo {title} {{Impurity and vortex States in the
  bilayer high-temperature superconductor La$_{3}$Ni$_{2}$O$_{7}$}},}\ }\href
  {\doibase 10.48550/arXiv.2308.07651} {\bibfield  {journal} {\bibinfo
  {journal} {arXiv preprint arXiv:2308.07651}\ } (\bibinfo {year} {2023}),\
  10.48550/arXiv.2308.07651}\BibitemShut {NoStop}%
\bibitem [{\citenamefont {Qin}\ and\ \citenamefont {Yang}(2023)}]{pair3}%
  \BibitemOpen
  \bibfield  {author} {\bibinfo {author} {\bibfnamefont {Qiong}\ \bibnamefont
  {Qin}}\ and\ \bibinfo {author} {\bibfnamefont {Yi-feng}\ \bibnamefont
  {Yang}},\ }\bibfield  {title} {\enquote {\bibinfo {title} {{High-$T_c$
  superconductivity by mobilizing local spin singlets and possible route to
  higher $T_c$ in pressurized La$_{3}$Ni$_{2}$O$_{7}$}},}\ }\href {\doibase
  10.48550/arXiv.2308.09044} {\bibfield  {journal} {\bibinfo  {journal} {arXiv
  preprint arXiv:2308.09044}\ } (\bibinfo {year} {2023}),\
  10.48550/arXiv.2308.09044}\BibitemShut {NoStop}%
\bibitem [{\citenamefont {Lu}\ \emph {et~al.}(2023)\citenamefont {Lu},
  \citenamefont {Pan}, \citenamefont {Yang},\ and\ \citenamefont {Wu}}]{pair4}%
  \BibitemOpen
  \bibfield  {author} {\bibinfo {author} {\bibfnamefont {Chen}\ \bibnamefont
  {Lu}}, \bibinfo {author} {\bibfnamefont {Zhiming}\ \bibnamefont {Pan}},
  \bibinfo {author} {\bibfnamefont {Fan}\ \bibnamefont {Yang}}, \ and\ \bibinfo
  {author} {\bibfnamefont {Congjun}\ \bibnamefont {Wu}},\ }\bibfield  {title}
  {\enquote {\bibinfo {title} {{Interlayer coupling driven high-temperature
  superconductivity in La$_3$Ni$_2$O$_{7}$ under pressure}},}\ }\href {\doibase
  10.48550/arXiv.2307.14965} {\bibfield  {journal} {\bibinfo  {journal} {arXiv
  preprint arXiv:2307.14965}\ } (\bibinfo {year} {2023}),\
  10.48550/arXiv.2307.14965}\BibitemShut {NoStop}%
\bibitem [{\citenamefont {Yang}\ \emph
  {et~al.}(2023{\natexlab{a}})\citenamefont {Yang}, \citenamefont {Liu},
  \citenamefont {Wang},\ and\ \citenamefont {Wang}}]{pair5}%
  \BibitemOpen
  \bibfield  {author} {\bibinfo {author} {\bibfnamefont {Qing-Geng}\
  \bibnamefont {Yang}}, \bibinfo {author} {\bibfnamefont {Han-Yang}\
  \bibnamefont {Liu}}, \bibinfo {author} {\bibfnamefont {Da}~\bibnamefont
  {Wang}}, \ and\ \bibinfo {author} {\bibfnamefont {Qiang-Hua}\ \bibnamefont
  {Wang}},\ }\bibfield  {title} {\enquote {\bibinfo {title} {{Possible
  $S_{\pm}$-wave superconductivity in La$_{3}$Ni$_{2}$O$_{7}$}},}\ }\href
  {\doibase 10.48550/arXiv.2306.03706} {\bibfield  {journal} {\bibinfo
  {journal} {arXiv preprint arXiv:2306.03706}\ } (\bibinfo {year}
  {2023}{\natexlab{a}}),\ 10.48550/arXiv.2306.03706}\BibitemShut {NoStop}%
\bibitem [{\citenamefont {Zhang}\ \emph
  {et~al.}(2023{\natexlab{c}})\citenamefont {Zhang}, \citenamefont {Lin},
  \citenamefont {Moreo}, \citenamefont {Maier},\ and\ \citenamefont
  {Dagotto}}]{pair6}%
  \BibitemOpen
  \bibfield  {author} {\bibinfo {author} {\bibfnamefont {Yang}\ \bibnamefont
  {Zhang}}, \bibinfo {author} {\bibfnamefont {Ling-Fang}\ \bibnamefont {Lin}},
  \bibinfo {author} {\bibfnamefont {Adriana}\ \bibnamefont {Moreo}}, \bibinfo
  {author} {\bibfnamefont {Thomas~A}\ \bibnamefont {Maier}}, \ and\ \bibinfo
  {author} {\bibfnamefont {Elbio}\ \bibnamefont {Dagotto}},\ }\bibfield
  {title} {\enquote {\bibinfo {title} {{Structural phase transition,
  $s_{\pm}$-wave pairing and magnetic stripe order in the bilayered nickelate
  superconductor La$_{3}$Ni$_{2}$O$_{7}$ under pressure}},}\ }\href {\doibase
  10.48550/arXiv.2307.15276} {\bibfield  {journal} {\bibinfo  {journal} {arXiv
  preprint arXiv:2307.15276}\ } (\bibinfo {year} {2023}{\natexlab{c}}),\
  10.48550/arXiv.2307.15276}\BibitemShut {NoStop}%
\bibitem [{\citenamefont {Liu}\ \emph {et~al.}(2023{\natexlab{a}})\citenamefont
  {Liu}, \citenamefont {Mei}, \citenamefont {Ye}, \citenamefont {Chen},\ and\
  \citenamefont {Yang}}]{pair7}%
  \BibitemOpen
  \bibfield  {author} {\bibinfo {author} {\bibfnamefont {Yu-Bo}\ \bibnamefont
  {Liu}}, \bibinfo {author} {\bibfnamefont {Jia-Wei}\ \bibnamefont {Mei}},
  \bibinfo {author} {\bibfnamefont {Fei}\ \bibnamefont {Ye}}, \bibinfo {author}
  {\bibfnamefont {Wei-Qiang}\ \bibnamefont {Chen}}, \ and\ \bibinfo {author}
  {\bibfnamefont {Fan}\ \bibnamefont {Yang}},\ }\bibfield  {title} {\enquote
  {\bibinfo {title} {{The $s_{\pm}$-Wave Pairing and the Destructive Role of
  Apical-Oxygen Deficiencies in La$_{3}$Ni$_{2}$O$_{7}$ Under Pressure}},}\
  }\href {\doibase 10.48550/arXiv.2307.10144} {\bibfield  {journal} {\bibinfo
  {journal} {arXiv preprint arXiv:2307.10144}\ } (\bibinfo {year}
  {2023}{\natexlab{a}}),\ 10.48550/arXiv.2307.10144}\BibitemShut {NoStop}%
\bibitem [{\citenamefont {Tian}\ \emph {et~al.}(2023)\citenamefont {Tian},
  \citenamefont {Chen}, \citenamefont {Wang}, \citenamefont {He},\ and\
  \citenamefont {Lu}}]{pair1}%
  \BibitemOpen
  \bibfield  {author} {\bibinfo {author} {\bibfnamefont {Yi-Heng}\ \bibnamefont
  {Tian}}, \bibinfo {author} {\bibfnamefont {Yin}\ \bibnamefont {Chen}},
  \bibinfo {author} {\bibfnamefont {Jia-Ming}\ \bibnamefont {Wang}}, \bibinfo
  {author} {\bibfnamefont {Rong-Qiang}\ \bibnamefont {He}}, \ and\ \bibinfo
  {author} {\bibfnamefont {Zhong-Yi}\ \bibnamefont {Lu}},\ }\bibfield  {title}
  {\enquote {\bibinfo {title} {{Correlation Effects and Concomitant Two-Orbital
  $s_{\pm}$-Wave Superconductivity in La$_{3}$Ni$_{2}$O$_{7}$ under High
  Pressure}},}\ }\href {\doibase 10.48550/arXiv.2308.09698} {\bibfield
  {journal} {\bibinfo  {journal} {arXiv preprint arXiv:2308.09698}\ } (\bibinfo
  {year} {2023}),\ 10.48550/arXiv.2308.09698}\BibitemShut {NoStop}%
\bibitem [{\citenamefont {Liao}\ \emph {et~al.}(2023)\citenamefont {Liao},
  \citenamefont {Chen}, \citenamefont {Duan}, \citenamefont {Wang},
  \citenamefont {Liu}, \citenamefont {Yu},\ and\ \citenamefont {Si}}]{pair2}%
  \BibitemOpen
  \bibfield  {author} {\bibinfo {author} {\bibfnamefont {Zhiguang}\
  \bibnamefont {Liao}}, \bibinfo {author} {\bibfnamefont {Lei}\ \bibnamefont
  {Chen}}, \bibinfo {author} {\bibfnamefont {Guijing}\ \bibnamefont {Duan}},
  \bibinfo {author} {\bibfnamefont {Yiming}\ \bibnamefont {Wang}}, \bibinfo
  {author} {\bibfnamefont {Changle}\ \bibnamefont {Liu}}, \bibinfo {author}
  {\bibfnamefont {Rong}\ \bibnamefont {Yu}}, \ and\ \bibinfo {author}
  {\bibfnamefont {Qimiao}\ \bibnamefont {Si}},\ }\bibfield  {title} {\enquote
  {\bibinfo {title} {{Electron correlations and superconductivity in
  La$_{3}$Ni$_{2}$O$_{7}$ under pressure tuning}},}\ }\href {\doibase
  10.48550/arXiv.2307.16697} {\bibfield  {journal} {\bibinfo  {journal} {arXiv
  preprint arXiv:2307.16697}\ } (\bibinfo {year} {2023}),\
  10.48550/arXiv.2307.16697}\BibitemShut {NoStop}%
\bibitem [{\citenamefont {Geisler}\ \emph {et~al.}(2024)\citenamefont
  {Geisler}, \citenamefont {Hamlin}, \citenamefont {Stewart}, \citenamefont
  {Hennig},\ and\ \citenamefont {Hirschfeld}}]{geisler2024structural}%
  \BibitemOpen
  \bibfield  {author} {\bibinfo {author} {\bibfnamefont {Benjamin}\
  \bibnamefont {Geisler}}, \bibinfo {author} {\bibfnamefont {James~J}\
  \bibnamefont {Hamlin}}, \bibinfo {author} {\bibfnamefont {Gregory~R}\
  \bibnamefont {Stewart}}, \bibinfo {author} {\bibfnamefont {Richard~G}\
  \bibnamefont {Hennig}}, \ and\ \bibinfo {author} {\bibfnamefont
  {PJ}~\bibnamefont {Hirschfeld}},\ }\bibfield  {title} {\enquote {\bibinfo
  {title} {{Structural transitions, octahedral rotations, and electronic
  properties of A$_{3}$Ni$_{2}$O$_{7}$ rare-earth nickelates under high
  pressure}},}\ }\href {\doibase 10.1038/s41535-024-00648-0} {\bibfield
  {journal} {\bibinfo  {journal} {npj Quantum Materials}\ }\textbf {\bibinfo
  {volume} {9}},\ \bibinfo {pages} {38} (\bibinfo {year} {2024})}\BibitemShut
  {NoStop}%
\bibitem [{\citenamefont {Liu}\ \emph {et~al.}(2024)\citenamefont {Liu},
  \citenamefont {Peng}, \citenamefont {Huang}, \citenamefont {Moritz},
  \citenamefont {Jia},\ and\ \citenamefont {Devereaux}}]{liu2024emergence}%
  \BibitemOpen
  \bibfield  {author} {\bibinfo {author} {\bibfnamefont {Fangze}\ \bibnamefont
  {Liu}}, \bibinfo {author} {\bibfnamefont {Cheng}\ \bibnamefont {Peng}},
  \bibinfo {author} {\bibfnamefont {Edwin~W}\ \bibnamefont {Huang}}, \bibinfo
  {author} {\bibfnamefont {Brian}\ \bibnamefont {Moritz}}, \bibinfo {author}
  {\bibfnamefont {Chunjing}\ \bibnamefont {Jia}}, \ and\ \bibinfo {author}
  {\bibfnamefont {Thomas~P}\ \bibnamefont {Devereaux}},\ }\bibfield  {title}
  {\enquote {\bibinfo {title} {Emergence of antiferromagnetic correlations and
  kondolike features in a model for infinite layer nickelates},}\ }\href
  {\doibase 10.1038/s41535-024-00659-x} {\bibfield  {journal} {\bibinfo
  {journal} {npj Quantum Materials}\ }\textbf {\bibinfo {volume} {9}},\
  \bibinfo {pages} {49} (\bibinfo {year} {2024})}\BibitemShut {NoStop}%
\bibitem [{\citenamefont {Liu}\ \emph {et~al.}(2023{\natexlab{b}})\citenamefont
  {Liu}, \citenamefont {Sun}, \citenamefont {Huo}, \citenamefont {Ma},
  \citenamefont {Ji}, \citenamefont {Yi}, \citenamefont {Li}, \citenamefont
  {Liu}, \citenamefont {Yu}, \citenamefont {Zhang} \emph
  {et~al.}}]{liu2023evidence}%
  \BibitemOpen
  \bibfield  {author} {\bibinfo {author} {\bibfnamefont {Zengjia}\ \bibnamefont
  {Liu}}, \bibinfo {author} {\bibfnamefont {Hualei}\ \bibnamefont {Sun}},
  \bibinfo {author} {\bibfnamefont {Mengwu}\ \bibnamefont {Huo}}, \bibinfo
  {author} {\bibfnamefont {Xiaoyan}\ \bibnamefont {Ma}}, \bibinfo {author}
  {\bibfnamefont {Yi}~\bibnamefont {Ji}}, \bibinfo {author} {\bibfnamefont
  {Enkui}\ \bibnamefont {Yi}}, \bibinfo {author} {\bibfnamefont {Lisi}\
  \bibnamefont {Li}}, \bibinfo {author} {\bibfnamefont {Hui}\ \bibnamefont
  {Liu}}, \bibinfo {author} {\bibfnamefont {Jia}\ \bibnamefont {Yu}}, \bibinfo
  {author} {\bibfnamefont {Ziyou}\ \bibnamefont {Zhang}},  \emph {et~al.},\
  }\bibfield  {title} {\enquote {\bibinfo {title} {{Evidence for charge and
  spin density waves in single crystals of La$_3$Ni$_2$O$_{7}$ and
  La$_3$Ni$_2$O$_{6}$}},}\ }\href {\doibase 10.1007/s11433-022-1962-4}
  {\bibfield  {journal} {\bibinfo  {journal} {Science China Physics, Mechanics
  \& Astronomy}\ }\textbf {\bibinfo {volume} {66}},\ \bibinfo {pages} {217411}
  (\bibinfo {year} {2023}{\natexlab{b}})}\BibitemShut {NoStop}%
\bibitem [{\citenamefont {Kakoi}\ \emph {et~al.}(2023)\citenamefont {Kakoi},
  \citenamefont {Oi}, \citenamefont {Ohshita}, \citenamefont {Yashima},
  \citenamefont {Kuroki}, \citenamefont {Adachi}, \citenamefont {Hatada},
  \citenamefont {Uda},\ and\ \citenamefont {Mukuda}}]{kakoi2023multiband}%
  \BibitemOpen
  \bibfield  {author} {\bibinfo {author} {\bibfnamefont {Masataka}\
  \bibnamefont {Kakoi}}, \bibinfo {author} {\bibfnamefont {Takashi}\
  \bibnamefont {Oi}}, \bibinfo {author} {\bibfnamefont {Yujiro}\ \bibnamefont
  {Ohshita}}, \bibinfo {author} {\bibfnamefont {Mitsuharu}\ \bibnamefont
  {Yashima}}, \bibinfo {author} {\bibfnamefont {Kazuhiko}\ \bibnamefont
  {Kuroki}}, \bibinfo {author} {\bibfnamefont {Yoshinobu}\ \bibnamefont
  {Adachi}}, \bibinfo {author} {\bibfnamefont {Naoyuki}\ \bibnamefont
  {Hatada}}, \bibinfo {author} {\bibfnamefont {Tetsuya}\ \bibnamefont {Uda}}, \
  and\ \bibinfo {author} {\bibfnamefont {Hidekazu}\ \bibnamefont {Mukuda}},\
  }\bibfield  {title} {\enquote {\bibinfo {title} {{Multiband Metallic Ground
  State in Multilayered Nickelates La$_3$Ni$_2$O$_7$ and La$_4$Ni$_3$O$_{10}$
  Revealed by $^{139}$La-NMR at Ambient Pressure}},}\ }\href {\doibase
  10.7566/JPSJ.93.053702} {\bibfield  {journal} {\bibinfo  {journal} {arXiv
  preprint arXiv:2312.11844}\ } (\bibinfo {year} {2023}),\
  10.7566/JPSJ.93.053702}\BibitemShut {NoStop}%
\bibitem [{\citenamefont {Chen}\ \emph
  {et~al.}(2024{\natexlab{a}})\citenamefont {Chen}, \citenamefont {Choi},
  \citenamefont {Jiang}, \citenamefont {Mei}, \citenamefont {Jiang},
  \citenamefont {Li}, \citenamefont {Agrestini}, \citenamefont
  {Garcia-Fernandez}, \citenamefont {Huang}, \citenamefont {Sun} \emph
  {et~al.}}]{chen2024electronic}%
  \BibitemOpen
  \bibfield  {author} {\bibinfo {author} {\bibfnamefont {Xiaoyang}\
  \bibnamefont {Chen}}, \bibinfo {author} {\bibfnamefont {Jaewon}\ \bibnamefont
  {Choi}}, \bibinfo {author} {\bibfnamefont {Zhicheng}\ \bibnamefont {Jiang}},
  \bibinfo {author} {\bibfnamefont {Jiong}\ \bibnamefont {Mei}}, \bibinfo
  {author} {\bibfnamefont {Kun}\ \bibnamefont {Jiang}}, \bibinfo {author}
  {\bibfnamefont {Jie}\ \bibnamefont {Li}}, \bibinfo {author} {\bibfnamefont
  {Stefano}\ \bibnamefont {Agrestini}}, \bibinfo {author} {\bibfnamefont
  {Mirian}\ \bibnamefont {Garcia-Fernandez}}, \bibinfo {author} {\bibfnamefont
  {Xing}\ \bibnamefont {Huang}}, \bibinfo {author} {\bibfnamefont {Hualei}\
  \bibnamefont {Sun}},  \emph {et~al.},\ }\bibfield  {title} {\enquote
  {\bibinfo {title} {{Electronic and magnetic excitations in
  La$_3$Ni$_2$O$_{7}$ }},}\ }\href {\doibase 10.48550/arXiv.2401.12657}
  {\bibfield  {journal} {\bibinfo  {journal} {arXiv preprint arXiv:2401.12657}\
  } (\bibinfo {year} {2024}{\natexlab{a}}),\
  10.48550/arXiv.2401.12657}\BibitemShut {NoStop}%
\bibitem [{\citenamefont {Dan}\ \emph {et~al.}(2024)\citenamefont {Dan},
  \citenamefont {Zhou}, \citenamefont {Huo}, \citenamefont {Wang},
  \citenamefont {Nie}, \citenamefont {Wang}, \citenamefont {Wu},\ and\
  \citenamefont {Chen}}]{dan2024spin}%
  \BibitemOpen
  \bibfield  {author} {\bibinfo {author} {\bibfnamefont {Zhao}\ \bibnamefont
  {Dan}}, \bibinfo {author} {\bibfnamefont {Yanbing}\ \bibnamefont {Zhou}},
  \bibinfo {author} {\bibfnamefont {Mengwu}\ \bibnamefont {Huo}}, \bibinfo
  {author} {\bibfnamefont {Yu}~\bibnamefont {Wang}}, \bibinfo {author}
  {\bibfnamefont {Linpeng}\ \bibnamefont {Nie}}, \bibinfo {author}
  {\bibfnamefont {Meng}\ \bibnamefont {Wang}}, \bibinfo {author} {\bibfnamefont
  {Tao}\ \bibnamefont {Wu}}, \ and\ \bibinfo {author} {\bibfnamefont {Xianhui}\
  \bibnamefont {Chen}},\ }\bibfield  {title} {\enquote {\bibinfo {title}
  {{Spin-density-wave transition in double-layer nickelate
  La$_3$Ni$_2$O$_{7}$}},}\ }\href {\doibase 10.48550/arXiv.2402.03952}
  {\bibfield  {journal} {\bibinfo  {journal} {arXiv preprint arXiv:2402.03952}\
  } (\bibinfo {year} {2024}),\ 10.48550/arXiv.2402.03952}\BibitemShut {NoStop}%
\bibitem [{\citenamefont {Xie}\ \emph {et~al.}(2024)\citenamefont {Xie},
  \citenamefont {Huo}, \citenamefont {Ni}, \citenamefont {Shen}, \citenamefont
  {Huang}, \citenamefont {Sun}, \citenamefont {Walker}, \citenamefont {Adroja},
  \citenamefont {Yu}, \citenamefont {Shen} \emph {et~al.}}]{xie2024neutron}%
  \BibitemOpen
  \bibfield  {author} {\bibinfo {author} {\bibfnamefont {Tao}\ \bibnamefont
  {Xie}}, \bibinfo {author} {\bibfnamefont {Mengwu}\ \bibnamefont {Huo}},
  \bibinfo {author} {\bibfnamefont {Xiaosheng}\ \bibnamefont {Ni}}, \bibinfo
  {author} {\bibfnamefont {Feiran}\ \bibnamefont {Shen}}, \bibinfo {author}
  {\bibfnamefont {Xing}\ \bibnamefont {Huang}}, \bibinfo {author}
  {\bibfnamefont {Hualei}\ \bibnamefont {Sun}}, \bibinfo {author}
  {\bibfnamefont {Helen~C}\ \bibnamefont {Walker}}, \bibinfo {author}
  {\bibfnamefont {Devashibhai}\ \bibnamefont {Adroja}}, \bibinfo {author}
  {\bibfnamefont {Dehong}\ \bibnamefont {Yu}}, \bibinfo {author} {\bibfnamefont
  {Bing}\ \bibnamefont {Shen}},  \emph {et~al.},\ }\bibfield  {title} {\enquote
  {\bibinfo {title} {{Strong interlayer magnetic exchange coupling in
  La$_3$Ni$_2$O$_{7-\delta}$ revealed by inelastic neutron scattering}},}\
  }\href {\doibase 10.1016/j.scib.2024.07.030} {\bibfield  {journal} {\bibinfo
  {journal} {Science Bulletin}\ } (\bibinfo {year} {2024}),\
  10.1016/j.scib.2024.07.030}\BibitemShut {NoStop}%
\bibitem [{\citenamefont {Chen}\ \emph
  {et~al.}(2024{\natexlab{b}})\citenamefont {Chen}, \citenamefont {Liu},
  \citenamefont {Jiao}, \citenamefont {Zou}, \citenamefont {Jiang},
  \citenamefont {Li}, \citenamefont {Luo}, \citenamefont {Wu}, \citenamefont
  {Zhang}, \citenamefont {Guo},\ and\ \citenamefont
  {Shu}}]{PhysRevLett.132.256503}%
  \BibitemOpen
  \bibfield  {author} {\bibinfo {author} {\bibfnamefont {Kaiwen}\ \bibnamefont
  {Chen}}, \bibinfo {author} {\bibfnamefont {Xiangqi}\ \bibnamefont {Liu}},
  \bibinfo {author} {\bibfnamefont {Jiachen}\ \bibnamefont {Jiao}}, \bibinfo
  {author} {\bibfnamefont {Muyuan}\ \bibnamefont {Zou}}, \bibinfo {author}
  {\bibfnamefont {Chengyu}\ \bibnamefont {Jiang}}, \bibinfo {author}
  {\bibfnamefont {Xin}\ \bibnamefont {Li}}, \bibinfo {author} {\bibfnamefont
  {Yixuan}\ \bibnamefont {Luo}}, \bibinfo {author} {\bibfnamefont {Qiong}\
  \bibnamefont {Wu}}, \bibinfo {author} {\bibfnamefont {Ningyuan}\ \bibnamefont
  {Zhang}}, \bibinfo {author} {\bibfnamefont {Yanfeng}\ \bibnamefont {Guo}}, \
  and\ \bibinfo {author} {\bibfnamefont {Lei}\ \bibnamefont {Shu}},\ }\bibfield
   {title} {\enquote {\bibinfo {title} {{Evidence of Spin Density Waves in
  La$_3$Ni$_2$O$_{7-\delta}$}},}\ }\href {\doibase
  10.1103/PhysRevLett.132.256503} {\bibfield  {journal} {\bibinfo  {journal}
  {Phys. Rev. Lett.}\ }\textbf {\bibinfo {volume} {132}},\ \bibinfo {pages}
  {256503} (\bibinfo {year} {2024}{\natexlab{b}})}\BibitemShut {NoStop}%
\bibitem [{\citenamefont {Yi}\ \emph {et~al.}(2024)\citenamefont {Yi},
  \citenamefont {Meng}, \citenamefont {Li}, \citenamefont {Liao}, \citenamefont
  {You}, \citenamefont {Gu},\ and\ \citenamefont
  {Su}}]{yi2024antiferromagnetic}%
  \BibitemOpen
  \bibfield  {author} {\bibinfo {author} {\bibfnamefont {Xin-Wei}\ \bibnamefont
  {Yi}}, \bibinfo {author} {\bibfnamefont {Ying}\ \bibnamefont {Meng}},
  \bibinfo {author} {\bibfnamefont {Jia-Wen}\ \bibnamefont {Li}}, \bibinfo
  {author} {\bibfnamefont {Zheng-Wei}\ \bibnamefont {Liao}}, \bibinfo {author}
  {\bibfnamefont {Jing-Yang}\ \bibnamefont {You}}, \bibinfo {author}
  {\bibfnamefont {Bo}~\bibnamefont {Gu}}, \ and\ \bibinfo {author}
  {\bibfnamefont {Gang}\ \bibnamefont {Su}},\ }\bibfield  {title} {\enquote
  {\bibinfo {title} {{Antiferromagnetic Ground State, Charge Density Waves and
  Oxygen Vacancies Induced Metal-Insulator Transition in Pressurized
  La$_3$Ni$_2$O$_{7}$}},}\ }\href {\doibase 10.48550/arXiv.2403.11455}
  {\bibfield  {journal} {\bibinfo  {journal} {arXiv preprint arXiv:2403.11455}\
  } (\bibinfo {year} {2024}),\ 10.48550/arXiv.2403.11455}\BibitemShut {NoStop}%
\bibitem [{\citenamefont {Zhang}\ \emph
  {et~al.}(2023{\natexlab{d}})\citenamefont {Zhang}, \citenamefont {Lin},
  \citenamefont {Moreo},\ and\ \citenamefont {Dagotto}}]{PhysRevB.108.L180510}%
  \BibitemOpen
  \bibfield  {author} {\bibinfo {author} {\bibfnamefont {Yang}\ \bibnamefont
  {Zhang}}, \bibinfo {author} {\bibfnamefont {Ling-Fang}\ \bibnamefont {Lin}},
  \bibinfo {author} {\bibfnamefont {Adriana}\ \bibnamefont {Moreo}}, \ and\
  \bibinfo {author} {\bibfnamefont {Elbio}\ \bibnamefont {Dagotto}},\
  }\bibfield  {title} {\enquote {\bibinfo {title} {{Electronic structure, dimer
  physics, orbital-selective behavior, and magnetic tendencies in the bilayer
  nickelate superconductor La$_3$Ni$_2$O$_{7}$ under pressure}},}\ }\href
  {\doibase 10.1103/PhysRevB.108.L180510} {\bibfield  {journal} {\bibinfo
  {journal} {Phys. Rev. B}\ }\textbf {\bibinfo {volume} {108}},\ \bibinfo
  {pages} {L180510} (\bibinfo {year} {2023}{\natexlab{d}})}\BibitemShut
  {NoStop}%
\bibitem [{\citenamefont {Zhang}\ \emph
  {et~al.}(2024{\natexlab{b}})\citenamefont {Zhang}, \citenamefont {Lin},
  \citenamefont {Moreo}, \citenamefont {Maier},\ and\ \citenamefont
  {Dagotto}}]{zhang2024structural}%
  \BibitemOpen
  \bibfield  {author} {\bibinfo {author} {\bibfnamefont {Yang}\ \bibnamefont
  {Zhang}}, \bibinfo {author} {\bibfnamefont {Ling-Fang}\ \bibnamefont {Lin}},
  \bibinfo {author} {\bibfnamefont {Adriana}\ \bibnamefont {Moreo}}, \bibinfo
  {author} {\bibfnamefont {Thomas~A}\ \bibnamefont {Maier}}, \ and\ \bibinfo
  {author} {\bibfnamefont {Elbio}\ \bibnamefont {Dagotto}},\ }\bibfield
  {title} {\enquote {\bibinfo {title} {{Structural phase transition,
  s$\pm$-wave pairing, and magnetic stripe order in bilayered superconductor
  La$_3$Ni$_2$O$_{7}$ under pressure}},}\ }\href {\doibase
  10.1038/s41467-024-46622-z} {\bibfield  {journal} {\bibinfo  {journal}
  {Nature Communications}\ }\textbf {\bibinfo {volume} {15}},\ \bibinfo {pages}
  {2470} (\bibinfo {year} {2024}{\natexlab{b}})}\BibitemShut {NoStop}%
\bibitem [{\citenamefont {Zhou}\ \emph {et~al.}(2024)\citenamefont {Zhou},
  \citenamefont {Chen}, \citenamefont {Choi}, \citenamefont {Jiang},
  \citenamefont {Mei}, \citenamefont {Jiang}, \citenamefont {Li}, \citenamefont
  {Agrestini}, \citenamefont {Garcia-Fernandez}, \citenamefont {Sun} \emph
  {et~al.}}]{zhou2024electronic}%
  \BibitemOpen
  \bibfield  {author} {\bibinfo {author} {\bibfnamefont {Ke-Jin}\ \bibnamefont
  {Zhou}}, \bibinfo {author} {\bibfnamefont {Xiaoyang}\ \bibnamefont {Chen}},
  \bibinfo {author} {\bibfnamefont {Jaewon}\ \bibnamefont {Choi}}, \bibinfo
  {author} {\bibfnamefont {Zhicheng}\ \bibnamefont {Jiang}}, \bibinfo {author}
  {\bibfnamefont {Jiong}\ \bibnamefont {Mei}}, \bibinfo {author} {\bibfnamefont
  {Kun}\ \bibnamefont {Jiang}}, \bibinfo {author} {\bibfnamefont {Jie}\
  \bibnamefont {Li}}, \bibinfo {author} {\bibfnamefont {S}~\bibnamefont
  {Agrestini}}, \bibinfo {author} {\bibfnamefont {Mirian}\ \bibnamefont
  {Garcia-Fernandez}}, \bibinfo {author} {\bibfnamefont {Hualei}\ \bibnamefont
  {Sun}},  \emph {et~al.},\ }\bibfield  {title} {\enquote {\bibinfo {title}
  {{Electronic and magnetic excitations in La$_3$Ni$_2$O$_7$}},}\ }\href
  {\doibase 10.48550/arXiv.2401.12657} {\  (\bibinfo {year} {2024}),\
  10.48550/arXiv.2401.12657}\BibitemShut {NoStop}%
\bibitem [{\citenamefont {Pardo}\ and\ \citenamefont
  {Pickett}(2011)}]{PhysRevB.83.245128}%
  \BibitemOpen
  \bibfield  {author} {\bibinfo {author} {\bibfnamefont {Victor}\ \bibnamefont
  {Pardo}}\ and\ \bibinfo {author} {\bibfnamefont {Warren~E.}\ \bibnamefont
  {Pickett}},\ }\bibfield  {title} {\enquote {\bibinfo {title}
  {{Metal-insulator transition in layered nickelates
  La${}_{3}$Ni${}_{2}$O${}_{7\ensuremath{-}\ensuremath{\delta}}$
  ($\ensuremath{\delta}$ $=$ 0.0, 0.5, 1)}},}\ }\href {\doibase
  10.1103/PhysRevB.83.245128} {\bibfield  {journal} {\bibinfo  {journal} {Phys.
  Rev. B}\ }\textbf {\bibinfo {volume} {83}},\ \bibinfo {pages} {245128}
  (\bibinfo {year} {2011})}\BibitemShut {NoStop}%
\bibitem [{\citenamefont {Sui}\ \emph {et~al.}(2024)\citenamefont {Sui},
  \citenamefont {Han}, \citenamefont {Jin}, \citenamefont {Chen}, \citenamefont
  {Qiao}, \citenamefont {Shao},\ and\ \citenamefont
  {Huang}}]{PhysRevB.109.205156}%
  \BibitemOpen
  \bibfield  {author} {\bibinfo {author} {\bibfnamefont {Xuelei}\ \bibnamefont
  {Sui}}, \bibinfo {author} {\bibfnamefont {Xiangru}\ \bibnamefont {Han}},
  \bibinfo {author} {\bibfnamefont {Heng}\ \bibnamefont {Jin}}, \bibinfo
  {author} {\bibfnamefont {Xiaojun}\ \bibnamefont {Chen}}, \bibinfo {author}
  {\bibfnamefont {Liang}\ \bibnamefont {Qiao}}, \bibinfo {author}
  {\bibfnamefont {Xiaohong}\ \bibnamefont {Shao}}, \ and\ \bibinfo {author}
  {\bibfnamefont {Bing}\ \bibnamefont {Huang}},\ }\bibfield  {title} {\enquote
  {\bibinfo {title} {{Electronic properties of the bilayer nickelates
  ${R}_{3}\mathrm{N}{\mathrm{i}}_{2}{\mathrm{O}}_{7}$ with oxygen vacancies
  ($R=\mathrm{La}$ or Ce)}},}\ }\href {\doibase 10.1103/PhysRevB.109.205156}
  {\bibfield  {journal} {\bibinfo  {journal} {Phys. Rev. B}\ }\textbf {\bibinfo
  {volume} {109}},\ \bibinfo {pages} {205156} (\bibinfo {year}
  {2024})}\BibitemShut {NoStop}%
\bibitem [{\citenamefont {Dudarev}\ \emph {et~al.}(1998)\citenamefont
  {Dudarev}, \citenamefont {Botton}, \citenamefont {Savrasov}, \citenamefont
  {Humphreys},\ and\ \citenamefont {Sutton}}]{dudarev1998electron}%
  \BibitemOpen
  \bibfield  {author} {\bibinfo {author} {\bibfnamefont {Sergei~L}\
  \bibnamefont {Dudarev}}, \bibinfo {author} {\bibfnamefont {Gianluigi~A}\
  \bibnamefont {Botton}}, \bibinfo {author} {\bibfnamefont {Sergey~Y}\
  \bibnamefont {Savrasov}}, \bibinfo {author} {\bibfnamefont {CJ}~\bibnamefont
  {Humphreys}}, \ and\ \bibinfo {author} {\bibfnamefont {Adrian~P}\
  \bibnamefont {Sutton}},\ }\bibfield  {title} {\enquote {\bibinfo {title}
  {{Electron-energy-loss spectra and the structural stability of nickel oxide:
  An LSDA + U study}},}\ }\href {\doibase 10.1103/PhysRevB.57.1505} {\bibfield
  {journal} {\bibinfo  {journal} {Physical Review B}\ }\textbf {\bibinfo
  {volume} {57}},\ \bibinfo {pages} {1505} (\bibinfo {year}
  {1998})}\BibitemShut {NoStop}%
\bibitem [{\citenamefont {Khasanov}\ \emph {et~al.}(2024)\citenamefont
  {Khasanov}, \citenamefont {Hicken}, \citenamefont {Gawryluk}, \citenamefont
  {Sorel}, \citenamefont {B{\"o}tzel}, \citenamefont {Lechermann},
  \citenamefont {Eremin}, \citenamefont {Luetkens},\ and\ \citenamefont
  {Guguchia}}]{khasanov2024pressure}%
  \BibitemOpen
  \bibfield  {author} {\bibinfo {author} {\bibfnamefont {Rustem}\ \bibnamefont
  {Khasanov}}, \bibinfo {author} {\bibfnamefont {Thomas~J}\ \bibnamefont
  {Hicken}}, \bibinfo {author} {\bibfnamefont {Dariusz~J}\ \bibnamefont
  {Gawryluk}}, \bibinfo {author} {\bibfnamefont {Lo{\"\i}c~Pierre}\
  \bibnamefont {Sorel}}, \bibinfo {author} {\bibfnamefont {Steffen}\
  \bibnamefont {B{\"o}tzel}}, \bibinfo {author} {\bibfnamefont {Frank}\
  \bibnamefont {Lechermann}}, \bibinfo {author} {\bibfnamefont {Ilya~M}\
  \bibnamefont {Eremin}}, \bibinfo {author} {\bibfnamefont {Hubertus}\
  \bibnamefont {Luetkens}}, \ and\ \bibinfo {author} {\bibfnamefont {Zurab}\
  \bibnamefont {Guguchia}},\ }\bibfield  {title} {\enquote {\bibinfo {title}
  {{Pressure-Induced Split of the Density Wave Transitions in
  La$_3$Ni$_{2}$O$_{7-\delta}$}},}\ }\href {\doibase 10.48550/arXiv.2402.10485}
  {\bibfield  {journal} {\bibinfo  {journal} {arXiv preprint arXiv:2402.10485}\
  } (\bibinfo {year} {2024}),\ 10.48550/arXiv.2402.10485}\BibitemShut {NoStop}%
\bibitem [{\citenamefont {LaBollita}\ \emph {et~al.}(2023)\citenamefont
  {LaBollita}, \citenamefont {Pardo}, \citenamefont {Norman},\ and\
  \citenamefont {Botana}}]{labollita2023electronic}%
  \BibitemOpen
  \bibfield  {author} {\bibinfo {author} {\bibfnamefont {Harrison}\
  \bibnamefont {LaBollita}}, \bibinfo {author} {\bibfnamefont {Victor}\
  \bibnamefont {Pardo}}, \bibinfo {author} {\bibfnamefont {Michael~R}\
  \bibnamefont {Norman}}, \ and\ \bibinfo {author} {\bibfnamefont {Antia~S}\
  \bibnamefont {Botana}},\ }\bibfield  {title} {\enquote {\bibinfo {title}
  {{Electronic structure and magnetic properties of La$_3$Ni$_2$O$_7$ under
  pressure}},}\ }\href {\doibase 10.48550/arXiv.2309.17279} {\bibfield
  {journal} {\bibinfo  {journal} {arXiv preprint arXiv:2309.17279}\ } (\bibinfo
  {year} {2023}),\ 10.48550/arXiv.2309.17279}\BibitemShut {NoStop}%
\bibitem [{\citenamefont {LaBollita}\ \emph {et~al.}(2024)\citenamefont
  {LaBollita}, \citenamefont {Pardo}, \citenamefont {Norman},\ and\
  \citenamefont {Botana}}]{labollita2024assessing}%
  \BibitemOpen
  \bibfield  {author} {\bibinfo {author} {\bibfnamefont {Harrison}\
  \bibnamefont {LaBollita}}, \bibinfo {author} {\bibfnamefont {Victor}\
  \bibnamefont {Pardo}}, \bibinfo {author} {\bibfnamefont {Michael~R}\
  \bibnamefont {Norman}}, \ and\ \bibinfo {author} {\bibfnamefont {Antia~S}\
  \bibnamefont {Botana}},\ }\bibfield  {title} {\enquote {\bibinfo {title}
  {{Assessing the formation of spin and charge stripes in La$_3$Ni$_2$O$_7$
  from first-principles}},}\ }\href {\doibase 10.48550/arXiv.2407.14409}
  {\bibfield  {journal} {\bibinfo  {journal} {arXiv preprint arXiv:2407.14409}\
  } (\bibinfo {year} {2024}),\ 10.48550/arXiv.2407.14409}\BibitemShut {NoStop}%
\bibitem [{\citenamefont {Zhang}\ \emph
  {et~al.}(2024{\natexlab{c}})\citenamefont {Zhang}, \citenamefont {Xu},\ and\
  \citenamefont {Xiang}}]{zhang2024emergent}%
  \BibitemOpen
  \bibfield  {author} {\bibinfo {author} {\bibfnamefont {Binhua}\ \bibnamefont
  {Zhang}}, \bibinfo {author} {\bibfnamefont {Changsong}\ \bibnamefont {Xu}}, \
  and\ \bibinfo {author} {\bibfnamefont {Hongjun}\ \bibnamefont {Xiang}},\
  }\bibfield  {title} {\enquote {\bibinfo {title} {{Emergent
  spin-charge-orbital order in superconductor La$_3$Ni$_2$O$_7$}},}\ }\href
  {\doibase 10.48550/arXiv.2407.18473} {\bibfield  {journal} {\bibinfo
  {journal} {arXiv preprint arXiv:2407.18473}\ } (\bibinfo {year}
  {2024}{\natexlab{c}}),\ 10.48550/arXiv.2407.18473}\BibitemShut {NoStop}%
\bibitem [{\citenamefont {Swendsen}\ and\ \citenamefont
  {Wang}(1986)}]{swendsen1986replica}%
  \BibitemOpen
  \bibfield  {author} {\bibinfo {author} {\bibfnamefont {Robert~H}\
  \bibnamefont {Swendsen}}\ and\ \bibinfo {author} {\bibfnamefont {Jian-Sheng}\
  \bibnamefont {Wang}},\ }\bibfield  {title} {\enquote {\bibinfo {title}
  {Replica monte carlo simulation of spin-glasses},}\ }\href {\doibase
  10.1103/PhysRevLett.57.2607} {\bibfield  {journal} {\bibinfo  {journal}
  {Physical review letters}\ }\textbf {\bibinfo {volume} {57}},\ \bibinfo
  {pages} {2607} (\bibinfo {year} {1986})}\BibitemShut {NoStop}%
\bibitem [{\citenamefont {Cao}\ \emph {et~al.}(2009)\citenamefont {Cao},
  \citenamefont {Guo}, \citenamefont {Vanderbilt},\ and\ \citenamefont
  {He}}]{mc}%
  \BibitemOpen
  \bibfield  {author} {\bibinfo {author} {\bibfnamefont {Kun}\ \bibnamefont
  {Cao}}, \bibinfo {author} {\bibfnamefont {Guang-Can}\ \bibnamefont {Guo}},
  \bibinfo {author} {\bibfnamefont {David}\ \bibnamefont {Vanderbilt}}, \ and\
  \bibinfo {author} {\bibfnamefont {Lixin}\ \bibnamefont {He}},\ }\bibfield
  {title} {\enquote {\bibinfo {title} {{First-Principles Modeling of
  Multiferroic $RMn_{2}O_{5}$}},}\ }\href {\doibase
  10.1103/PhysRevLett.103.257201} {\bibfield  {journal} {\bibinfo  {journal}
  {Phys. Rev. Lett.}\ }\textbf {\bibinfo {volume} {103}},\ \bibinfo {pages}
  {257201} (\bibinfo {year} {2009})}\BibitemShut {NoStop}%
\bibitem [{\citenamefont {Dong}\ \emph {et~al.}(2024)\citenamefont {Dong},
  \citenamefont {Huo}, \citenamefont {Li}, \citenamefont {Li}, \citenamefont
  {Li}, \citenamefont {Sun}, \citenamefont {Gu}, \citenamefont {Lu},
  \citenamefont {Wang}, \citenamefont {Wang} \emph
  {et~al.}}]{dong2023visualization}%
  \BibitemOpen
  \bibfield  {author} {\bibinfo {author} {\bibfnamefont {Zehao}\ \bibnamefont
  {Dong}}, \bibinfo {author} {\bibfnamefont {Mengwu}\ \bibnamefont {Huo}},
  \bibinfo {author} {\bibfnamefont {Jie}\ \bibnamefont {Li}}, \bibinfo {author}
  {\bibfnamefont {Jingyuan}\ \bibnamefont {Li}}, \bibinfo {author}
  {\bibfnamefont {Pengcheng}\ \bibnamefont {Li}}, \bibinfo {author}
  {\bibfnamefont {Hualei}\ \bibnamefont {Sun}}, \bibinfo {author}
  {\bibfnamefont {Lin}\ \bibnamefont {Gu}}, \bibinfo {author} {\bibfnamefont
  {Yi}~\bibnamefont {Lu}}, \bibinfo {author} {\bibfnamefont {Meng}\
  \bibnamefont {Wang}}, \bibinfo {author} {\bibfnamefont {Yayu}\ \bibnamefont
  {Wang}},  \emph {et~al.},\ }\bibfield  {title} {\enquote {\bibinfo {title}
  {{Visualization of Oxygen Vacancies and Self-doped Ligand Holes
  inLa$_3$Ni$_2$O$_{7-\delta}$}},}\ }\href {\doibase
  10.1038/s41586-024-07482-1} {\bibfield  {journal} {\bibinfo  {journal}
  {Nature}\ ,\ \bibinfo {pages} {1--6}} (\bibinfo {year} {2024})}\BibitemShut
  {NoStop}%
\bibitem [{\citenamefont {Gao}\ \emph {et~al.}(2024)\citenamefont {Gao},
  \citenamefont {Jin}, \citenamefont {Huyan}, \citenamefont {Ni}, \citenamefont
  {Wang}, \citenamefont {Xu}, \citenamefont {Bud’ko}, \citenamefont
  {Canfield}, \citenamefont {Xie},\ and\ \citenamefont
  {Cava}}]{gao2024la3ni2o6}%
  \BibitemOpen
  \bibfield  {author} {\bibinfo {author} {\bibfnamefont {Ran}\ \bibnamefont
  {Gao}}, \bibinfo {author} {\bibfnamefont {Lun}\ \bibnamefont {Jin}}, \bibinfo
  {author} {\bibfnamefont {Shuyuan}\ \bibnamefont {Huyan}}, \bibinfo {author}
  {\bibfnamefont {Danrui}\ \bibnamefont {Ni}}, \bibinfo {author} {\bibfnamefont
  {Haozhe}\ \bibnamefont {Wang}}, \bibinfo {author} {\bibfnamefont {Xianghan}\
  \bibnamefont {Xu}}, \bibinfo {author} {\bibfnamefont {Sergey~L}\ \bibnamefont
  {Bud’ko}}, \bibinfo {author} {\bibfnamefont {Paul}\ \bibnamefont
  {Canfield}}, \bibinfo {author} {\bibfnamefont {Weiwei}\ \bibnamefont {Xie}},
  \ and\ \bibinfo {author} {\bibfnamefont {Robert~J}\ \bibnamefont {Cava}},\
  }\bibfield  {title} {\enquote {\bibinfo {title} {{Is La$_3$Ni$_2$O$_{6.5}$ a
  Bulk Superconducting Nickelate?}}}\ }\href {\doibase 10.1021/acsami.3c17376}
  {\bibfield  {journal} {\bibinfo  {journal} {ACS Applied Materials \&
  Interfaces}\ } (\bibinfo {year} {2024}),\ 10.1021/acsami.3c17376}\BibitemShut
  {NoStop}%
\bibitem [{\citenamefont {Maity}\ \emph {et~al.}(2023)\citenamefont {Maity},
  \citenamefont {Ceretti}, \citenamefont {Keller}, \citenamefont {Schefer},
  \citenamefont {Roessli}, \citenamefont {Stuhr}, \citenamefont {Niedermayer},\
  and\ \citenamefont {Paulus}}]{PhysRevMaterials.7.024412}%
  \BibitemOpen
  \bibfield  {author} {\bibinfo {author} {\bibfnamefont {Sumit~Ranjan}\
  \bibnamefont {Maity}}, \bibinfo {author} {\bibfnamefont {Monica}\
  \bibnamefont {Ceretti}}, \bibinfo {author} {\bibfnamefont {Lukas}\
  \bibnamefont {Keller}}, \bibinfo {author} {\bibfnamefont {J\"urg}\
  \bibnamefont {Schefer}}, \bibinfo {author} {\bibfnamefont {Bertrand}\
  \bibnamefont {Roessli}}, \bibinfo {author} {\bibfnamefont {Uwe}\ \bibnamefont
  {Stuhr}}, \bibinfo {author} {\bibfnamefont {Christof}\ \bibnamefont
  {Niedermayer}}, \ and\ \bibinfo {author} {\bibfnamefont {Werner}\
  \bibnamefont {Paulus}},\ }\bibfield  {title} {\enquote {\bibinfo {title}
  {Evidence of correlated incommensurate structural and magnetic order in
  highly oxygen-doped layered nickelate
  ${\mathrm{nd}}_{2}{\mathrm{nio}}_{4.23}$},}\ }\href {\doibase
  10.1103/PhysRevMaterials.7.024412} {\bibfield  {journal} {\bibinfo  {journal}
  {Phys. Rev. Mater.}\ }\textbf {\bibinfo {volume} {7}},\ \bibinfo {pages}
  {024412} (\bibinfo {year} {2023})}\BibitemShut {NoStop}%
\bibitem [{\citenamefont {Dutta}\ \emph {et~al.}(2020)\citenamefont {Dutta},
  \citenamefont {Maity}, \citenamefont {Marsicano}, \citenamefont {Stewart},
  \citenamefont {Opel},\ and\ \citenamefont {Paulus}}]{PhysRevB.102.165130}%
  \BibitemOpen
  \bibfield  {author} {\bibinfo {author} {\bibfnamefont {Rajesh}\ \bibnamefont
  {Dutta}}, \bibinfo {author} {\bibfnamefont {Avishek}\ \bibnamefont {Maity}},
  \bibinfo {author} {\bibfnamefont {Anna}\ \bibnamefont {Marsicano}}, \bibinfo
  {author} {\bibfnamefont {J.~Ross}\ \bibnamefont {Stewart}}, \bibinfo {author}
  {\bibfnamefont {Matthias}\ \bibnamefont {Opel}}, \ and\ \bibinfo {author}
  {\bibfnamefont {Werner}\ \bibnamefont {Paulus}},\ }\bibfield  {title}
  {\enquote {\bibinfo {title} {Direct evidence for anisotropic
  three-dimensional magnetic excitations in a hole-doped antiferromagnet},}\
  }\href {\doibase 10.1103/PhysRevB.102.165130} {\bibfield  {journal} {\bibinfo
   {journal} {Phys. Rev. B}\ }\textbf {\bibinfo {volume} {102}},\ \bibinfo
  {pages} {165130} (\bibinfo {year} {2020})}\BibitemShut {NoStop}%
\bibitem [{\citenamefont {Hou}\ \emph {et~al.}(2023{\natexlab{b}})\citenamefont
  {Hou}, \citenamefont {Yang}, \citenamefont {Liu}, \citenamefont {Li},
  \citenamefont {Shan}, \citenamefont {Ma}, \citenamefont {Wang}, \citenamefont
  {Wang}, \citenamefont {Guo}, \citenamefont {Sun} \emph
  {et~al.}}]{hou2023emergence}%
  \BibitemOpen
  \bibfield  {author} {\bibinfo {author} {\bibfnamefont {Jun}\ \bibnamefont
  {Hou}}, \bibinfo {author} {\bibfnamefont {Peng-Tao}\ \bibnamefont {Yang}},
  \bibinfo {author} {\bibfnamefont {Zi-Yi}\ \bibnamefont {Liu}}, \bibinfo
  {author} {\bibfnamefont {Jing-Yuan}\ \bibnamefont {Li}}, \bibinfo {author}
  {\bibfnamefont {Peng-Fei}\ \bibnamefont {Shan}}, \bibinfo {author}
  {\bibfnamefont {Liang}\ \bibnamefont {Ma}}, \bibinfo {author} {\bibfnamefont
  {Gang}\ \bibnamefont {Wang}}, \bibinfo {author} {\bibfnamefont {Ning-Ning}\
  \bibnamefont {Wang}}, \bibinfo {author} {\bibfnamefont {Hai-Zhong}\
  \bibnamefont {Guo}}, \bibinfo {author} {\bibfnamefont {Jian-Ping}\
  \bibnamefont {Sun}},  \emph {et~al.},\ }\bibfield  {title} {\enquote
  {\bibinfo {title} {Emergence of high-temperature superconducting phase in
  pressurized la3ni2o7 crystals},}\ }\href {\doibase
  10.1088/0256-307X/40/11/117302} {\bibfield  {journal} {\bibinfo  {journal}
  {Chinese Physics Letters}\ }\textbf {\bibinfo {volume} {40}},\ \bibinfo
  {pages} {117302} (\bibinfo {year} {2023}{\natexlab{b}})}\BibitemShut
  {NoStop}%
\bibitem [{\citenamefont {Fan}\ \emph {et~al.}(2024)\citenamefont {Fan},
  \citenamefont {Luo}, \citenamefont {Huo}, \citenamefont {Wang}, \citenamefont
  {Li}, \citenamefont {Yang}, \citenamefont {Wang}, \citenamefont {Yao},\ and\
  \citenamefont {Wen}}]{PhysRevB.110.134520}%
  \BibitemOpen
  \bibfield  {author} {\bibinfo {author} {\bibfnamefont {Shengtai}\
  \bibnamefont {Fan}}, \bibinfo {author} {\bibfnamefont {Zhihui}\ \bibnamefont
  {Luo}}, \bibinfo {author} {\bibfnamefont {Mengwu}\ \bibnamefont {Huo}},
  \bibinfo {author} {\bibfnamefont {Zhaohui}\ \bibnamefont {Wang}}, \bibinfo
  {author} {\bibfnamefont {Han}\ \bibnamefont {Li}}, \bibinfo {author}
  {\bibfnamefont {Huan}\ \bibnamefont {Yang}}, \bibinfo {author} {\bibfnamefont
  {Meng}\ \bibnamefont {Wang}}, \bibinfo {author} {\bibfnamefont {Dao-Xin}\
  \bibnamefont {Yao}}, \ and\ \bibinfo {author} {\bibfnamefont {Hai-Hu}\
  \bibnamefont {Wen}},\ }\bibfield  {title} {\enquote {\bibinfo {title}
  {{Tunneling spectra with gaplike features observed in nickelate
  La$_3$Ni$_2$O$_7$ at ambient pressure}},}\ }\href {\doibase
  10.1103/PhysRevB.110.134520} {\bibfield  {journal} {\bibinfo  {journal}
  {Phys. Rev. B}\ }\textbf {\bibinfo {volume} {110}},\ \bibinfo {pages}
  {134520} (\bibinfo {year} {2024})}\BibitemShut {NoStop}%
\bibitem [{\citenamefont {Dai}(2015)}]{dai2015antiferromagnetic}%
  \BibitemOpen
  \bibfield  {author} {\bibinfo {author} {\bibfnamefont {Pengcheng}\
  \bibnamefont {Dai}},\ }\bibfield  {title} {\enquote {\bibinfo {title}
  {Antiferromagnetic order and spin dynamics in iron-based superconductors},}\
  }\href {\doibase 10.1103/RevModPhys.87.855} {\bibfield  {journal} {\bibinfo
  {journal} {Reviews of Modern Physics}\ }\textbf {\bibinfo {volume} {87}},\
  \bibinfo {pages} {855--896} (\bibinfo {year} {2015})}\BibitemShut {NoStop}%
\bibitem [{\citenamefont {Fang}\ \emph {et~al.}(2009)\citenamefont {Fang},
  \citenamefont {Bernevig},\ and\ \citenamefont {Hu}}]{fang2009theory}%
  \BibitemOpen
  \bibfield  {author} {\bibinfo {author} {\bibfnamefont {Chen}\ \bibnamefont
  {Fang}}, \bibinfo {author} {\bibfnamefont {B~Andrei}\ \bibnamefont
  {Bernevig}}, \ and\ \bibinfo {author} {\bibfnamefont {Jiangping}\
  \bibnamefont {Hu}},\ }\bibfield  {title} {\enquote {\bibinfo {title} {{Theory
  of magnetic order in Fe$_{1+y}$Te$_{1-x}$Se$_{x}$}},}\ }\href {\doibase
  10.1209/0295-5075/86/67005/meta} {\bibfield  {journal} {\bibinfo  {journal}
  {Europhysics Letters}\ }\textbf {\bibinfo {volume} {86}},\ \bibinfo {pages}
  {67005} (\bibinfo {year} {2009})}\BibitemShut {NoStop}%
\bibitem [{\citenamefont {Subedi}\ \emph {et~al.}(2008)\citenamefont {Subedi},
  \citenamefont {Zhang}, \citenamefont {Singh},\ and\ \citenamefont
  {Du}}]{subedi2008density}%
  \BibitemOpen
  \bibfield  {author} {\bibinfo {author} {\bibfnamefont {Alaska}\ \bibnamefont
  {Subedi}}, \bibinfo {author} {\bibfnamefont {Lijun}\ \bibnamefont {Zhang}},
  \bibinfo {author} {\bibfnamefont {David~J}\ \bibnamefont {Singh}}, \ and\
  \bibinfo {author} {\bibfnamefont {Mao-Hua}\ \bibnamefont {Du}},\ }\bibfield
  {title} {\enquote {\bibinfo {title} {{Density functional study of FeS, FeSe,
  and FeTe: Electronic structure, magnetism, phonons, and
  superconductivity}},}\ }\href {\doibase 10.1103/PhysRevB.78.134514}
  {\bibfield  {journal} {\bibinfo  {journal} {Physical Review B—Condensed
  Matter and Materials Physics}\ }\textbf {\bibinfo {volume} {78}},\ \bibinfo
  {pages} {134514} (\bibinfo {year} {2008})}\BibitemShut {NoStop}%
\bibitem [{\citenamefont {Toth}\ and\ \citenamefont
  {Lake}(2015)}]{toth2015linear}%
  \BibitemOpen
  \bibfield  {author} {\bibinfo {author} {\bibfnamefont {S}~\bibnamefont
  {Toth}}\ and\ \bibinfo {author} {\bibfnamefont {B}~\bibnamefont {Lake}},\
  }\bibfield  {title} {\enquote {\bibinfo {title} {{Linear spin wave theory for
  single-Q incommensurate magnetic structures}},}\ }\href {\doibase
  10.1088/0953-8984/27/16/166002} {\bibfield  {journal} {\bibinfo  {journal}
  {Journal of Physics: Condensed Matter}\ }\textbf {\bibinfo {volume} {27}},\
  \bibinfo {pages} {166002} (\bibinfo {year} {2015})}\BibitemShut {NoStop}%
\bibitem [{\citenamefont {Wu}\ \emph {et~al.}(2021)\citenamefont {Wu},
  \citenamefont {Shen}, \citenamefont {Barbour}, \citenamefont {Wang},
  \citenamefont {Prabhakaran}, \citenamefont {Boothroyd}, \citenamefont
  {Mazzoli}, \citenamefont {Tranquada}, \citenamefont {Dean},\ and\
  \citenamefont {Robinson}}]{PhysRevLett.127.275301}%
  \BibitemOpen
  \bibfield  {author} {\bibinfo {author} {\bibfnamefont {Longlong}\
  \bibnamefont {Wu}}, \bibinfo {author} {\bibfnamefont {Yao}\ \bibnamefont
  {Shen}}, \bibinfo {author} {\bibfnamefont {Andi~M.}\ \bibnamefont {Barbour}},
  \bibinfo {author} {\bibfnamefont {Wei}\ \bibnamefont {Wang}}, \bibinfo
  {author} {\bibfnamefont {Dharmalingam}\ \bibnamefont {Prabhakaran}}, \bibinfo
  {author} {\bibfnamefont {Andrew~T.}\ \bibnamefont {Boothroyd}}, \bibinfo
  {author} {\bibfnamefont {Claudio}\ \bibnamefont {Mazzoli}}, \bibinfo {author}
  {\bibfnamefont {John~M.}\ \bibnamefont {Tranquada}}, \bibinfo {author}
  {\bibfnamefont {Mark P.~M.}\ \bibnamefont {Dean}}, \ and\ \bibinfo {author}
  {\bibfnamefont {Ian~K.}\ \bibnamefont {Robinson}},\ }\bibfield  {title}
  {\enquote {\bibinfo {title} {Real space imaging of spin stripe domain
  fluctuations in a complex oxide},}\ }\href {\doibase
  10.1103/PhysRevLett.127.275301} {\bibfield  {journal} {\bibinfo  {journal}
  {Phys. Rev. Lett.}\ }\textbf {\bibinfo {volume} {127}},\ \bibinfo {pages}
  {275301} (\bibinfo {year} {2021})}\BibitemShut {NoStop}%
\bibitem [{\citenamefont {Dutta}\ \emph {et~al.}(2022)\citenamefont {Dutta},
  \citenamefont {Maity}, \citenamefont {Marsicano}, \citenamefont {Stewart},
  \citenamefont {Opel},\ and\ \citenamefont {Paulus}}]{PhysRevB.105.195147}%
  \BibitemOpen
  \bibfield  {author} {\bibinfo {author} {\bibfnamefont {Rajesh}\ \bibnamefont
  {Dutta}}, \bibinfo {author} {\bibfnamefont {Avishek}\ \bibnamefont {Maity}},
  \bibinfo {author} {\bibfnamefont {Anna}\ \bibnamefont {Marsicano}}, \bibinfo
  {author} {\bibfnamefont {J.~Ross}\ \bibnamefont {Stewart}}, \bibinfo {author}
  {\bibfnamefont {Matthias}\ \bibnamefont {Opel}}, \ and\ \bibinfo {author}
  {\bibfnamefont {Werner}\ \bibnamefont {Paulus}},\ }\bibfield  {title}
  {\enquote {\bibinfo {title} {{Revealing the effect of interstitial oxygen on
  the low-energy crystal electric field excitations of ${\mathrm{Pr}}^{3+}$ in
  $\mathit{214}$-nickelates}},}\ }\href {\doibase 10.1103/PhysRevB.105.195147}
  {\bibfield  {journal} {\bibinfo  {journal} {Phys. Rev. B}\ }\textbf {\bibinfo
  {volume} {105}},\ \bibinfo {pages} {195147} (\bibinfo {year}
  {2022})}\BibitemShut {NoStop}%
\bibitem [{\citenamefont {Kresse}\ and\ \citenamefont
  {Hafner}(1993)}]{kresse1993ab}%
  \BibitemOpen
  \bibfield  {author} {\bibinfo {author} {\bibfnamefont {G.}~\bibnamefont
  {Kresse}}\ and\ \bibinfo {author} {\bibfnamefont {J.}~\bibnamefont
  {Hafner}},\ }\bibfield  {title} {\enquote {\bibinfo {title} {{Ab initio
  molecular dynamics for open-shell transition metals}},}\ }\href {\doibase
  10.1103/PhysRevB.48.13115} {\bibfield  {journal} {\bibinfo  {journal} {Phys.
  Rev. B}\ }\textbf {\bibinfo {volume} {48}},\ \bibinfo {pages} {13115--13118}
  (\bibinfo {year} {1993})}\BibitemShut {NoStop}%
\bibitem [{kre(1996)}]{kresse1996efficient}%
  \BibitemOpen
  \bibfield  {title} {\enquote {\bibinfo {title} {{Efficient iterative schemes
  for ab initio total-energy calculations using a plane-wave basis set}, author
  = {Kresse, G. and Furthm\"uller, J.}}}\ }\href {\doibase
  10.1103/PhysRevB.54.11169} {\bibfield  {journal} {\bibinfo  {journal} {Phys.
  Rev. B}\ }\textbf {\bibinfo {volume} {54}},\ \bibinfo {pages} {11169--11186}
  (\bibinfo {year} {1996})}\BibitemShut {NoStop}%
\bibitem [{\citenamefont {Bl\"ochl}(1994)}]{blochl1994}%
  \BibitemOpen
  \bibfield  {author} {\bibinfo {author} {\bibfnamefont {P.~E.}\ \bibnamefont
  {Bl\"ochl}},\ }\bibfield  {title} {\enquote {\bibinfo {title} {{Projector
  augmented-wave method}},}\ }\href {\doibase 10.1103/PhysRevB.50.17953}
  {\bibfield  {journal} {\bibinfo  {journal} {Phys. Rev. B}\ }\textbf {\bibinfo
  {volume} {50}},\ \bibinfo {pages} {17953--17979} (\bibinfo {year}
  {1994})}\BibitemShut {NoStop}%
\bibitem [{\citenamefont {Yang}\ \emph
  {et~al.}(2023{\natexlab{b}})\citenamefont {Yang}, \citenamefont {Sun},
  \citenamefont {Hu}, \citenamefont {Xie}, \citenamefont {Miao}, \citenamefont
  {Luo}, \citenamefont {Chen}, \citenamefont {Liang}, \citenamefont {Zhu},
  \citenamefont {Qu} \emph {et~al.}}]{29}%
  \BibitemOpen
  \bibfield  {author} {\bibinfo {author} {\bibfnamefont {Jiangang}\
  \bibnamefont {Yang}}, \bibinfo {author} {\bibfnamefont {Hualei}\ \bibnamefont
  {Sun}}, \bibinfo {author} {\bibfnamefont {Xunwu}\ \bibnamefont {Hu}},
  \bibinfo {author} {\bibfnamefont {Yuyang}\ \bibnamefont {Xie}}, \bibinfo
  {author} {\bibfnamefont {Taimin}\ \bibnamefont {Miao}}, \bibinfo {author}
  {\bibfnamefont {Hailan}\ \bibnamefont {Luo}}, \bibinfo {author}
  {\bibfnamefont {Hao}\ \bibnamefont {Chen}}, \bibinfo {author} {\bibfnamefont
  {Bo}~\bibnamefont {Liang}}, \bibinfo {author} {\bibfnamefont {Wenpei}\
  \bibnamefont {Zhu}}, \bibinfo {author} {\bibfnamefont {Gexing}\ \bibnamefont
  {Qu}},  \emph {et~al.},\ }\bibfield  {title} {\enquote {\bibinfo {title}
  {{Orbital-Dependent Electron Correlation in Double-Layer Nickelate
  La$_3$Ni$_2$O$_7$}},}\ }\href {\doibase 10.48550/arXiv.2309.01148} {\bibfield
   {journal} {\bibinfo  {journal} {arXiv preprint arXiv:2309.01148}\ }
  (\bibinfo {year} {2023}{\natexlab{b}}),\
  10.48550/arXiv.2309.01148}\BibitemShut {NoStop}%
\bibitem [{\citenamefont {{Heyd,Jochen and Scuseria,Gustavo E. and
  Ernzerhof,Matthias }}(2003)}]{heyd2003hybrid}%
  \BibitemOpen
  \bibfield  {author} {\bibinfo {author} {\bibnamefont {{Heyd,Jochen and
  Scuseria,Gustavo E. and Ernzerhof,Matthias }}},\ }\bibfield  {title}
  {\enquote {\bibinfo {title} {Hybrid functionals based on a screened coulomb
  potential},}\ }\href {\doibase 10.1063/1.1564060} {\bibfield  {journal}
  {\bibinfo  {journal} {J. Chem. Phys.}\ }\textbf {\bibinfo {volume} {118}},\
  \bibinfo {pages} {8207--8215} (\bibinfo {year} {2003})}\BibitemShut {NoStop}%
\bibitem [{\citenamefont {Krukau}\ \emph {et~al.}(2006)\citenamefont {Krukau},
  \citenamefont {Vydrov}, \citenamefont {Izmaylov},\ and\ \citenamefont
  {Scuseria}}]{krukau2006influence}%
  \BibitemOpen
  \bibfield  {author} {\bibinfo {author} {\bibfnamefont {Aliaksandr~V.}\
  \bibnamefont {Krukau}}, \bibinfo {author} {\bibfnamefont {Oleg~A.}\
  \bibnamefont {Vydrov}}, \bibinfo {author} {\bibfnamefont {Artur~F.}\
  \bibnamefont {Izmaylov}}, \ and\ \bibinfo {author} {\bibfnamefont
  {Gustavo~E.}\ \bibnamefont {Scuseria}},\ }\bibfield  {title} {\enquote
  {\bibinfo {title} {{Influence of the exchange screening parameter on the
  performance of screened hybrid functionals}},}\ }\href {\doibase
  10.1063/1.2404663} {\bibfield  {journal} {\bibinfo  {journal} {J. Chem.
  Phys.}\ }\textbf {\bibinfo {volume} {125}},\ \bibinfo {pages} {224106}
  (\bibinfo {year} {2006})}\BibitemShut {NoStop}%
\bibitem [{\citenamefont {Heyd}\ \emph {et~al.}(2006)\citenamefont {Heyd},
  \citenamefont {Scuseria},\ and\ \citenamefont {Ernzerhof}}]{heyd2006}%
  \BibitemOpen
  \bibfield  {author} {\bibinfo {author} {\bibfnamefont {Jochen}\ \bibnamefont
  {Heyd}}, \bibinfo {author} {\bibfnamefont {Gustavo~E.}\ \bibnamefont
  {Scuseria}}, \ and\ \bibinfo {author} {\bibfnamefont {Matthias}\ \bibnamefont
  {Ernzerhof}},\ }\bibfield  {title} {\enquote {\bibinfo {title} {{Erratum:
  “Hybrid functionals based on a screened Coulomb potential” [J. Chem.
  Phys. 118, 8207 (2003)]}},}\ }\href {\doibase 10.1063/1.2204597} {\bibfield
  {journal} {\bibinfo  {journal} {J. Chem. Phys.}\ }\textbf {\bibinfo {volume}
  {124}},\ \bibinfo {pages} {219906} (\bibinfo {year} {2006})}\BibitemShut
  {NoStop}%
\bibitem [{\citenamefont {Lin}\ \emph {et~al.}(2020)\citenamefont {Lin},
  \citenamefont {Ren},\ and\ \citenamefont {He}}]{lin2020accuracy}%
  \BibitemOpen
  \bibfield  {author} {\bibinfo {author} {\bibfnamefont {Peize}\ \bibnamefont
  {Lin}}, \bibinfo {author} {\bibfnamefont {Xinguo}\ \bibnamefont {Ren}}, \
  and\ \bibinfo {author} {\bibfnamefont {Lixin}\ \bibnamefont {He}},\
  }\bibfield  {title} {\enquote {\bibinfo {title} {{Accuracy of Localized
  Resolution of the Identity in Periodic Hybrid Functional Calculations with
  Numerical Atomic Orbitals}},}\ }\href {\doibase 10.1021/acs.jpclett.0c00481}
  {\bibfield  {journal} {\bibinfo  {journal} {J. Phys. Chem. Lett.}\ }\textbf
  {\bibinfo {volume} {11}},\ \bibinfo {pages} {3082--3088} (\bibinfo {year}
  {2020})}\BibitemShut {NoStop}%
\bibitem [{\citenamefont {Lin}\ \emph {et~al.}(2021{\natexlab{a}})\citenamefont
  {Lin}, \citenamefont {Ren},\ and\ \citenamefont {He}}]{Lin2021}%
  \BibitemOpen
  \bibfield  {author} {\bibinfo {author} {\bibfnamefont {Peize}\ \bibnamefont
  {Lin}}, \bibinfo {author} {\bibfnamefont {Xinguo}\ \bibnamefont {Ren}}, \
  and\ \bibinfo {author} {\bibfnamefont {Lixin}\ \bibnamefont {He}},\
  }\bibfield  {title} {\enquote {\bibinfo {title} {{Efficient Hybrid Density
  Functional Calculations for Large Periodic Systems Using Numerical Atomic
  Orbitals}},}\ }\href {\doibase 10.1021/acs.jctc.0c00960} {\bibfield
  {journal} {\bibinfo  {journal} {J. Chem. Theory Comput.}\ }\textbf {\bibinfo
  {volume} {17}},\ \bibinfo {pages} {222--239} (\bibinfo {year}
  {2021}{\natexlab{a}})}\BibitemShut {NoStop}%
\bibitem [{\citenamefont {Ji}\ \emph {et~al.}(2022)\citenamefont {Ji},
  \citenamefont {Lin}, \citenamefont {Ren},\ and\ \citenamefont {He}}]{ji2022}%
  \BibitemOpen
  \bibfield  {author} {\bibinfo {author} {\bibfnamefont {Yuyang}\ \bibnamefont
  {Ji}}, \bibinfo {author} {\bibfnamefont {Peize}\ \bibnamefont {Lin}},
  \bibinfo {author} {\bibfnamefont {Xinguo}\ \bibnamefont {Ren}}, \ and\
  \bibinfo {author} {\bibfnamefont {Lixin}\ \bibnamefont {He}},\ }\bibfield
  {title} {\enquote {\bibinfo {title} {{Reproducibility of Hybrid Density
  Functional Calculations for Equation-of-State Properties and Band Gaps}},}\
  }\href {\doibase 10.1021/acs.jpca.2c05170} {\bibfield  {journal} {\bibinfo
  {journal} {J. Phys. Chem. A}\ }\textbf {\bibinfo {volume} {126}},\ \bibinfo
  {pages} {5924--5931} (\bibinfo {year} {2022})}\BibitemShut {NoStop}%
\bibitem [{\citenamefont {Chen}\ \emph {et~al.}(2010)\citenamefont {Chen},
  \citenamefont {Guo},\ and\ \citenamefont {He}}]{chen2010systematically}%
  \BibitemOpen
  \bibfield  {author} {\bibinfo {author} {\bibfnamefont {Mohan}\ \bibnamefont
  {Chen}}, \bibinfo {author} {\bibfnamefont {G-C}\ \bibnamefont {Guo}}, \ and\
  \bibinfo {author} {\bibfnamefont {Lixin}\ \bibnamefont {He}},\ }\bibfield
  {title} {\enquote {\bibinfo {title} {{Systematically improvable optimized
  atomic basis sets for ab initio calculations}},}\ }\href {\doibase
  10.1088/0953-8984/22/44/445501} {\bibfield  {journal} {\bibinfo  {journal}
  {J. Phys. Condens. Matter}\ }\textbf {\bibinfo {volume} {22}},\ \bibinfo
  {pages} {445501} (\bibinfo {year} {2010})}\BibitemShut {NoStop}%
\bibitem [{\citenamefont {Li}\ \emph {et~al.}(2016)\citenamefont {Li},
  \citenamefont {Liu}, \citenamefont {Chen}, \citenamefont {Lin}, \citenamefont
  {Ren}, \citenamefont {Lin}, \citenamefont {Yang},\ and\ \citenamefont
  {He}}]{li2016large}%
  \BibitemOpen
  \bibfield  {author} {\bibinfo {author} {\bibfnamefont {Pengfei}\ \bibnamefont
  {Li}}, \bibinfo {author} {\bibfnamefont {Xiaohui}\ \bibnamefont {Liu}},
  \bibinfo {author} {\bibfnamefont {Mohan}\ \bibnamefont {Chen}}, \bibinfo
  {author} {\bibfnamefont {Peize}\ \bibnamefont {Lin}}, \bibinfo {author}
  {\bibfnamefont {Xinguo}\ \bibnamefont {Ren}}, \bibinfo {author}
  {\bibfnamefont {Lin}\ \bibnamefont {Lin}}, \bibinfo {author} {\bibfnamefont
  {Chao}\ \bibnamefont {Yang}}, \ and\ \bibinfo {author} {\bibfnamefont
  {Lixin}\ \bibnamefont {He}},\ }\bibfield  {title} {\enquote {\bibinfo {title}
  {{Large-scale ab initio simulations based on systematically improvable atomic
  basis}},}\ }\href {\doibase 10.1016/j.commatsci.2015.07.004} {\bibfield
  {journal} {\bibinfo  {journal} {Comput. Mater. Sci.}\ }\textbf {\bibinfo
  {volume} {112}},\ \bibinfo {pages} {503--517} (\bibinfo {year} {2016})},\
  \bibinfo {note} {computational Materials Science in China}\BibitemShut
  {NoStop}%
\bibitem [{\citenamefont {Vanderbilt}(1990)}]{vanderbilt1990soft}%
  \BibitemOpen
  \bibfield  {author} {\bibinfo {author} {\bibfnamefont {David}\ \bibnamefont
  {Vanderbilt}},\ }\bibfield  {title} {\enquote {\bibinfo {title} {{Soft
  self-consistent pseudopotentials in a generalized eigenvalue formalism}},}\
  }\href {\doibase 10.1103/PhysRevB.41.7892} {\bibfield  {journal} {\bibinfo
  {journal} {Phys. Rev. B}\ }\textbf {\bibinfo {volume} {41}},\ \bibinfo
  {pages} {7892} (\bibinfo {year} {1990})}\BibitemShut {NoStop}%
\bibitem [{\citenamefont {Schlipf}\ and\ \citenamefont
  {Gygi}(2015)}]{schlipf2015optimization}%
  \BibitemOpen
  \bibfield  {author} {\bibinfo {author} {\bibfnamefont {Martin}\ \bibnamefont
  {Schlipf}}\ and\ \bibinfo {author} {\bibfnamefont {Fran{\c{c}}ois}\
  \bibnamefont {Gygi}},\ }\bibfield  {title} {\enquote {\bibinfo {title}
  {{Optimization algorithm for the generation of ONCV pseudopotentials}},}\
  }\href {\doibase 10.1016/j.cpc.2015.05.011} {\bibfield  {journal} {\bibinfo
  {journal} {Comput Phys Commun}\ }\textbf {\bibinfo {volume} {196}},\ \bibinfo
  {pages} {36--44} (\bibinfo {year} {2015})}\BibitemShut {NoStop}%
\bibitem [{\citenamefont {Lin}\ \emph {et~al.}(2021{\natexlab{b}})\citenamefont
  {Lin}, \citenamefont {Ren},\ and\ \citenamefont {He}}]{lin2021strategy}%
  \BibitemOpen
  \bibfield  {author} {\bibinfo {author} {\bibfnamefont {Peize}\ \bibnamefont
  {Lin}}, \bibinfo {author} {\bibfnamefont {Xinguo}\ \bibnamefont {Ren}}, \
  and\ \bibinfo {author} {\bibfnamefont {Lixin}\ \bibnamefont {He}},\
  }\bibfield  {title} {\enquote {\bibinfo {title} {{Strategy for constructing
  compact numerical atomic orbital basis sets by incorporating the gradients of
  reference wavefunctions}},}\ }\href {\doibase 10.1103/PhysRevB.103.235131}
  {\bibfield  {journal} {\bibinfo  {journal} {Phys. Rev. B}\ }\textbf {\bibinfo
  {volume} {103}},\ \bibinfo {pages} {235131} (\bibinfo {year}
  {2021}{\natexlab{b}})}\BibitemShut {NoStop}%
\bibitem [{\citenamefont {Pizzi}\ \emph {et~al.}(2020)\citenamefont {Pizzi},
  \citenamefont {Vitale}, \citenamefont {Arita}, \citenamefont {Blügel},
  \citenamefont {Freimuth}, \citenamefont {Géranton}, \citenamefont
  {Gibertini}, \citenamefont {Gresch}, \citenamefont {Johnson}, \citenamefont
  {Koretsune}, \citenamefont {Ibañez-Azpiroz}, \citenamefont {Lee},
  \citenamefont {Lihm}, \citenamefont {Marchand}, \citenamefont {Marrazzo},
  \citenamefont {Mokrousov}, \citenamefont {Mustafa}, \citenamefont {Nohara},
  \citenamefont {Nomura}, \citenamefont {Paulatto}, \citenamefont {Poncé},
  \citenamefont {Ponweiser}, \citenamefont {Qiao}, \citenamefont {Thöle},
  \citenamefont {Tsirkin}, \citenamefont {Wierzbowska}, \citenamefont
  {Marzari}, \citenamefont {Vanderbilt}, \citenamefont {Souza}, \citenamefont
  {Mostofi},\ and\ \citenamefont {Yates}}]{Pizzi_2020}%
  \BibitemOpen
  \bibfield  {author} {\bibinfo {author} {\bibfnamefont {Giovanni}\
  \bibnamefont {Pizzi}}, \bibinfo {author} {\bibfnamefont {Valerio}\
  \bibnamefont {Vitale}}, \bibinfo {author} {\bibfnamefont {Ryotaro}\
  \bibnamefont {Arita}}, \bibinfo {author} {\bibfnamefont {Stefan}\
  \bibnamefont {Blügel}}, \bibinfo {author} {\bibfnamefont {Frank}\
  \bibnamefont {Freimuth}}, \bibinfo {author} {\bibfnamefont {Guillaume}\
  \bibnamefont {Géranton}}, \bibinfo {author} {\bibfnamefont {Marco}\
  \bibnamefont {Gibertini}}, \bibinfo {author} {\bibfnamefont {Dominik}\
  \bibnamefont {Gresch}}, \bibinfo {author} {\bibfnamefont {Charles}\
  \bibnamefont {Johnson}}, \bibinfo {author} {\bibfnamefont {Takashi}\
  \bibnamefont {Koretsune}}, \bibinfo {author} {\bibfnamefont {Julen}\
  \bibnamefont {Ibañez-Azpiroz}}, \bibinfo {author} {\bibfnamefont {Hyungjun}\
  \bibnamefont {Lee}}, \bibinfo {author} {\bibfnamefont {Jae-Mo}\ \bibnamefont
  {Lihm}}, \bibinfo {author} {\bibfnamefont {Daniel}\ \bibnamefont {Marchand}},
  \bibinfo {author} {\bibfnamefont {Antimo}\ \bibnamefont {Marrazzo}}, \bibinfo
  {author} {\bibfnamefont {Yuriy}\ \bibnamefont {Mokrousov}}, \bibinfo {author}
  {\bibfnamefont {Jamal~I}\ \bibnamefont {Mustafa}}, \bibinfo {author}
  {\bibfnamefont {Yoshiro}\ \bibnamefont {Nohara}}, \bibinfo {author}
  {\bibfnamefont {Yusuke}\ \bibnamefont {Nomura}}, \bibinfo {author}
  {\bibfnamefont {Lorenzo}\ \bibnamefont {Paulatto}}, \bibinfo {author}
  {\bibfnamefont {Samuel}\ \bibnamefont {Poncé}}, \bibinfo {author}
  {\bibfnamefont {Thomas}\ \bibnamefont {Ponweiser}}, \bibinfo {author}
  {\bibfnamefont {Junfeng}\ \bibnamefont {Qiao}}, \bibinfo {author}
  {\bibfnamefont {Florian}\ \bibnamefont {Thöle}}, \bibinfo {author}
  {\bibfnamefont {Stepan~S}\ \bibnamefont {Tsirkin}}, \bibinfo {author}
  {\bibfnamefont {Małgorzata}\ \bibnamefont {Wierzbowska}}, \bibinfo {author}
  {\bibfnamefont {Nicola}\ \bibnamefont {Marzari}}, \bibinfo {author}
  {\bibfnamefont {David}\ \bibnamefont {Vanderbilt}}, \bibinfo {author}
  {\bibfnamefont {Ivo}\ \bibnamefont {Souza}}, \bibinfo {author} {\bibfnamefont
  {Arash~A}\ \bibnamefont {Mostofi}}, \ and\ \bibinfo {author} {\bibfnamefont
  {Jonathan~R}\ \bibnamefont {Yates}},\ }\bibfield  {title} {\enquote {\bibinfo
  {title} {Wannier90 as a community code: new features and applications},}\
  }\href {\doibase 10.1088/1361-648X/ab51ff} {\ \textbf {\bibinfo {volume}
  {32}},\ \bibinfo {pages} {165902} (\bibinfo {year} {2020})}\BibitemShut
  {NoStop}%
\bibitem [{HE2(2021)}]{HE2021107938}%
  \BibitemOpen
  \bibfield  {title} {\enquote {\bibinfo {title} {{TB2J: A python package for
  computing magnetic interaction parameters}},}\ }\href {\doibase
  https://doi.org/10.1016/j.cpc.2021.107938} {\bibfield  {journal} {\bibinfo
  {journal} {Computer Physics Communications}\ }\textbf {\bibinfo {volume}
  {264}},\ \bibinfo {pages} {107938} (\bibinfo {year} {2021})}\BibitemShut
  {NoStop}%
\bibitem [{\citenamefont {Chen}\ \emph {et~al.}(2023)\citenamefont {Chen},
  \citenamefont {Liu}, \citenamefont {Jiao}, \citenamefont {Zou}, \citenamefont
  {Luo}, \citenamefont {Wu}, \citenamefont {Zhang}, \citenamefont {Guo},\ and\
  \citenamefont {Shu}}]{chen2023evidence}%
  \BibitemOpen
  \bibfield  {author} {\bibinfo {author} {\bibfnamefont {Kaiwen}\ \bibnamefont
  {Chen}}, \bibinfo {author} {\bibfnamefont {Xiangqi}\ \bibnamefont {Liu}},
  \bibinfo {author} {\bibfnamefont {Jiachen}\ \bibnamefont {Jiao}}, \bibinfo
  {author} {\bibfnamefont {Myyuan}\ \bibnamefont {Zou}}, \bibinfo {author}
  {\bibfnamefont {Yixuan}\ \bibnamefont {Luo}}, \bibinfo {author}
  {\bibfnamefont {Qiong}\ \bibnamefont {Wu}}, \bibinfo {author} {\bibfnamefont
  {Ningyuan}\ \bibnamefont {Zhang}}, \bibinfo {author} {\bibfnamefont
  {Yanfeng}\ \bibnamefont {Guo}}, \ and\ \bibinfo {author} {\bibfnamefont
  {Lei}\ \bibnamefont {Shu}},\ }\bibfield  {title} {\enquote {\bibinfo {title}
  {{Evidence of spin density waves in La$_3$Ni$_2$O$_{7-\delta}$ }},}\ }\href
  {\doibase 10.48550/arXiv.2311.15717} {\bibfield  {journal} {\bibinfo
  {journal} {arXiv preprint arXiv:2311.15717}\ } (\bibinfo {year} {2023}),\
  10.48550/arXiv.2311.15717}\BibitemShut {NoStop}%
\end{thebibliography}

~\\
\noindent{\bf\large Methods}

\noindent{\bf DFT + $U$ and HSE06 calculations.} First-principles calculations are carried out with the Vienna $ab\ initio$ Simulation Package (VASP)~\cite{kresse1993ab, kresse1996efficient}. We use the Perdew Burke-Ernzerhof functional with a spin-polarized generalized gradient approximation (GGA). The projector augmented-wave (PAW)~\cite{blochl1994} method with a 550 eV plane wave cutoff is employed. The spin-polarized GGA is combined with onsite Coulomb interactions~\cite{dudarev1998electron} included for Ni 3$d$ orbitals (GGA + $U$)~\cite{20,29}. The scheme is implemented using effective on-site Hubbard $U$ parameter $U_{\text{eff}}$ = $U - J$, where $J$ is fixed to $J$ = 1.0 eV. 
For HSE06~\cite{heyd2003hybrid,krukau2006influence,heyd2006,lin2020accuracy,Lin2021,ji2022} calculations, we employ the atomic-orbital-based {\it ab initio} computation at UStc (ABACUS) code package \cite{chen2010systematically, li2016large}, which makes use of the SG15 optimized norm-conserving Vanderbilt-type (ONCV) pseudopotentials \cite{vanderbilt1990soft, schlipf2015optimization} and the so-called ``DZP-DPSI'' \cite{lin2021strategy} numerical atomic orbital basis sets. \\

\noindent{\bf Heisenberg exchange interactions and MC calculations.} Utilizing the Wannier90 code~\cite{Pizzi_2020}, we derive a tight-binding model employing maximally localized Wannier functions associated with La $d$ $f$ orbitals, Ni $d$ orbitals, and O $s$ $p$ orbitals. Subsequently, the TB2J package~\cite{HE2021107938} is employed as a postprocessing tool to compute the Heisenberg exchange interactions based on the magnetic force theorem. 
 
To construct the tight-binding model, our calculated DOS reveals that the $d$ and $f$ orbitals of La, the $d$ orbitals of Ni, and the $s$ and $p$ orbitals of O contribute significantly around the Fermi level (see Fig.~S1 in Supplementary Note 2 of the supplementary materials). Therefore, the projection block includes these orbitals, defining a set of localized functions used to generate an initial guess for the unitary transformations. A broad frozen energy window spanning $-6$ to $4$ eV was employed to accurately capture the higher energy conduction bands, with the projection centers serving as reference points during the Wannierisation procedure.  The final Wannier-interpolated band structures are in good agreement with  DFT-calculated band structures. (see Fig.~S2 in Supplementary Note 2 of the supplementary materials)

The magnetic phase diagrams are obtained using MC simulations. Due to the complexity of frustrated interactions, traditional serial-temperature MC methods are often insufficient for sampling. Therefore, we employed the replica-exchange MC method~\cite{swendsen1986replica,mc}, where multiple replicas are simulated concurrently across a range of temperatures, enabling configurational exchanges between them. This approach allows higher-temperature replicas to facilitate broad phase space exploration for the lower-temperature replicas, helping to avoid trapping in local energy minima. In our simulations, each MC sweep includes attempts over all variable states. The MC simulations for the cases with zero vacancy and ordered vacancies are performed with a 16 $\times$ 16 $\times$ 2 superlattice with 1 000 000 MC steps for statistics at each temperature. We performed the simulations over a temperature range from 1 to 300 K, adjusting temperatures to maintain an exchange rate of approximately 20\% between neighboring replicas. For each temperature, an initial 100 sweeps were conducted to equilibrate the system.
\\

\noindent{\bf Oxygen vacancies simulation.} Oxygen vacancies are simulated by removing the corresponding number of oxygen atoms from a La$_3$Ni$_2$O$_7$ supercell. For magnetic moment calculations, we employ a 96-atom supercell. Experimentally synthesized La$_3$Ni$_2$O$_7$ samples are commonly found to contain oxygen vacancies, with $\delta$ typically ranging from 0 to 0.5~\cite{gao2024la3ni2o6,dong2023visualization,chen2023evidence,dan2024spin,khasanov2024pressure} Therefore, to reduce computational cost from the supercell approach, exchange interactions with oxygen vacancies are computed only for $\delta = 0.25$ and $\delta = 0.5$ using a 48-atom supercell, where removal of one oxygen atom yields $\delta = 0.25$. In direct DFT calculations, a symmetric distribution of vacancies is adopted to mimic vacancies in real samples. \\

\noindent{\bf MC calculations with disordered oxygen vacancies.}  To further explore the magnetic phase diagrams in La$_3$Ni$_2$O$_7$, we conduct additional calculations using the calculated exchange interactions $J$'s and MC simulations~\cite{swendsen1986replica,mc}. Given that the calculated $J$'s indicate weak interactions between different bilayers, we simplify the calculations by focusing on a single bilayer. We first construct an 8×8 bilayer structure containing 128 Ni atoms. To simulate the disorder of oxygen vacancies, we randomly select a certain number of Ni pairs to be non-magnetic (charge). For instance, for $\delta = 0.25$, we randomly selected 16 out of 64 Ni pairs to be non-magnetic. We then set all $J$'s directly involving charge sites to zero. We perform simulations for a range of $\delta$ values, with selected simulated $T_{SDW}$ and magnetic ground states shown in Supplementary Note 6 FIG. S8. In the MC simulations with disordered vacancies, we use a bilayer with 32 $\times$ 32 $\times$ 1 superlattice, which contains a total of 2048 Ni atoms.\\

\noindent{\bf Magnetic excitation spectra simulations with a random splitting of $J$'s.}  To better simulate the effects of random oxygen vacancies, we group the original 12 $J$’s values into respective pools and randomly select bonds from each pool, assigning these values to the 48 $J$'s (derived from the original 12 $J$'s). For example, $J_2^{0.25}$ to $J_5^{0.25}$ are derived from $J_1$ in La$_3$Ni$_2$O$_7$, with their coordination numbers all being 4 in a supercell containing 32 Ni atoms. We randomly selected four bonds as $J_2^{0.25}$, another four as $J_3^{0.25}$, and so forth, resulting in a reclassified set of $J_2^{0.25}$ to $J_5^{0.25}$. For computational convenience, we focus on the major interactions, redistributing $J_1$ $\sim$ $J_7$ and $J_{12}$. This approach results in a random splitting of the $J$’s values, better reflecting the characteristics of random oxygen vacancies. 
Through quench optimization of the magnetic structure, we determine the magnetic ground state in the random scenario using a 2 $\times$ 2 $\times$ 1 supercell and obtain the corresponding theoretical spin wave.
\\

~\\
\noindent {\bf\large Data availability}\\
All data needed to evaluate the conclusions in the paper are available within the article. All raw data generated during the current study are available from the corresponding author upon request.

~\\
\noindent {\bf\large Code availability}\\
The codes used for the DFT calculations in this study are
available from the corresponding authors upon  request.

\vspace{3mm}

\noindent {\bf\large Acknowledgements}\\
Work at SYSU was supported by the National Key Research and Development Program of China (Grant Nos. 2023YFA1406500, 2022YFA1402802), the National Natural Science Foundation of China (Grant Nos.12174454, 12304187, 92165204),  the Guangdong Basic and Applied Basic Research Funds (Grant Nos. 2024B1515020040, 2022A1515011618), Guangzhou Basic and Applied Basic Research Funds (Grant Nos. 2024A04J6417, 2024A04J4024), and Guangdong Provincial Key Laboratory of Magnetoelectric Physics and Devices (Grant No. 2022B1212010008), Fundamental Research Funds for the Central Universities, 
Sun Yat-sen University (Grants No. 23ptpy158), Leading Talent Program of Guangdong Special Projects (201626003) and the open research fund of Songshan Lake Materials Laboratory (Grant no. 2023SLABFN30). Work at USTC is supported by Chinese National Science Foundation Grant No.12134012.

\vspace{3mm}

\noindent {\bf\large Author Contributions}\\
X.S.N. and K.C. proposed and designed the research. X.S.N. and K.C. contributed to the DFT and MC calculations. Y.Y.J. and L.X.H. contributed to HSE06 calculation. X.S.N., T.X, M.W. and K.C. contributed to the Magnetic excitation spectra simulations. X.S.N., T.X, D.X.Y., M.W. and K.C. analyzed the data. X.S.N. and K.C. wrote the paper. All authors participated in the discussion and comment on the paper.

\vspace{3mm}

\noindent{\bf\large Competing interests}\\
The authors declare no competing interests.

\end{document}